\documentclass[sigconf]{acmart}
\pdfoutput=1
\usepackage{epsfig,endnotes}
\usepackage{color}
\usepackage{xspace}
\usepackage{url}
\usepackage{multirow}
\usepackage{makecell}
\usepackage{listings}
\usepackage{pifont}
\usepackage[]{algorithm2e}
\usepackage{comment}
\usepackage{hyperref}
\usepackage{tabularx}
\usepackage{flushend}

\newcolumntype{Y}{>{\centering\arraybackslash}X}
\hypersetup{
    colorlinks = true,
    linkcolor = black,
    anchorcolor = black,
    citecolor = black,
    filecolor = black,
    urlcolor = black
}

\setcopyright{none} 

\begin{document}

\newcommand{\bheading}[1]{{\vspace{4pt}\noindent{\textbf{#1}}}}
\newcommand{\iheading}[1]{{\vspace{4pt}\noindent{\textit{#1}}}}

\newcommand{\etal}{\emph{et al.}\xspace}
\newcommand{\etc}{\emph{etc}\xspace}
\newcommand{\ie}{\emph{i.e.}\xspace}
\newcommand{\eg}{\emph{e.g.}\xspace}

\newcommand{\figurewidth}{\columnwidth}
\newcommand{\secref}[1]{\mbox{Sec.~\ref{#1}}\xspace}
\newcommand{\secrefs}[2]{\mbox{Sec.~\ref{#1}--\ref{#2}}\xspace}
\newcommand{\figref}[1]{\mbox{Fig.~\ref{#1}}}
\newcommand{\tabref}[1]{\mbox{Table~\ref{#1}}}
\newcommand{\lstref}[1]{\mbox{Listing~\ref{#1}}}
\newcommand{\appref}[1]{\mbox{Appendix~\ref{#1}}}
\newcommand{\ignore}[1]{}

\newcommand\st[1]{\textbf{Step \ding{#1}}}

\newcommand{\attack}{{{\sc SgxPectre}}\xspace}
\newcommand{\attackNames}{{{\sc SgxPectre} Attacks}\xspace}
\newcommand{\attackName}{{{\sc SgxPectre} Attack}\xspace}

\newcommand{\flushreload}{\textsc{Flush-Reload}\xspace}
\newcommand{\Flush}{\textsc{Flush}\xspace}
\newcommand{\Reload}{\textsc{Reload}\xspace}
\newcommand{\primeprobe}{\textsc{Prime-Probe}\xspace}
\newcommand{\Prime}{\textsc{Prime}\xspace}
\newcommand{\Probe}{\textsc{Probe}\xspace}
\newcommand{\evictreload}{\textsc{Evict-Reload}\xspace}
\newcommand{\flushflush}{\textsc{Flush-Flush}\xspace}
\newcommand{\flushfunc}{\texttt{clearcache}\xspace}

\newcommand{\gbytes}{\ensuremath{\mathrm{GB}}\xspace}
\newcommand{\mbytes}{\ensuremath{\mathrm{MB}}\xspace}
\newcommand{\kbytes}{\ensuremath{\mathrm{KB}}\xspace}
\newcommand{\bytes}{\ensuremath{\mathrm{B}}\xspace}
\newcommand{\hertz}{\ensuremath{\mathrm{Hz}}\xspace}
\newcommand{\ghertz}{\ensuremath{\mathrm{GHz}}\xspace}
\newcommand{\msecs}{\ensuremath{\mathrm{ms}}\xspace}
\newcommand{\usecs}{\ensuremath{\mathrm{\mu{}s}}\xspace}
\newcommand{\nsecs}{\ensuremath{\mathrm{ns}}\xspace}
\newcommand{\secs}{\ensuremath{\mathrm{s}}\xspace}
\newcommand{\gbits}{\ensuremath{\mathrm{Gb}}\xspace}

\newcommand\yz[1]{\textcolor{red}{\{\textbf{yinqian:} {\em#1}\}}}
\newcommand\gx[1]{\textcolor{green}{\{\textbf{guoxing:} {\em#1}\}}}
\newcommand\zq[1]{\textcolor{blue}{\{\textbf{Zhiqiang:} {\em#1}\}}}
\newcommand\scc[1]{\textcolor[rgb]{0.858, 0.188, 0.478}{\{\textbf{sanchuan:} {\em#1}\}}}

\newcounter{packednmbr}
\newenvironment{packedenumerate}{
\begin{list}{\thepackednmbr.}{\usecounter{packednmbr}
\setlength{\itemsep}{0pt}
\addtolength{\labelwidth}{4pt}
\setlength{\leftmargin}{12pt}
\setlength{\listparindent}{\parindent}
\setlength{\parsep}{3pt}
\setlength{\topsep}{3pt}}}{\end{list}}

\newenvironment{packeditemize}{
\begin{list}{$\bullet$}{
\setlength{\labelwidth}{8pt}
\setlength{\itemsep}{0pt}
\setlength{\leftmargin}{\labelwidth}
\addtolength{\leftmargin}{\labelsep}
\setlength{\parindent}{0pt}
\setlength{\listparindent}{\parindent}
\setlength{\parsep}{2pt}
\setlength{\topsep}{1pt}}}{\end{list}}

\newcommand{\cmark}{\ding{51}}%
\newcommand{\xmark}{\ding{55}}%

\lstdefinelanguage
   [x64]{Assembler}     
   [x86masm]{Assembler} 
   {morekeywords={CDQE,CQO,CMPSQ,CMPXCHG16B,JRCXZ,LODSQ,MOVSXD, %
                  POPFQ,PUSHFQ,SCASQ,STOSQ,IRETQ,RDTSCP,SWAPGS, %
                  rax,rdx,rcx,rbx,rsi,rdi,rsp,rbp, %
                  r8,r8d,r8w,r8b,r9,r9d,r9w,r9b,r10,r11,r12,enclu,mfence,lfence,cmova,cmovg,cmovl}}
                  
\definecolor{auburn}{rgb}{0.43, 0.21, 0.1}
\definecolor{codegreen}{rgb}{0,0.6,0}
\definecolor{codegray}{rgb}{0.5,0.5,0.5}
\definecolor{codepurple}{rgb}{0.58,0,0.82}
 
\lstdefinestyle{mystyle}{
	language=[x64]Assembler,
    backgroundcolor=\color{white},   
    commentstyle=\color{blue},
    keywordstyle=\color{black},
    numberstyle=\tiny\color{black},
    stringstyle=\color{black},
    basicstyle=\scriptsize,
    breakatwhitespace=false,         
    breaklines=true,                 
    captionpos=b,                    
    keepspaces=true,                 
    numbers=left,                    
    numbersep=5pt,                  
    showspaces=false,
    xleftmargin=0.25cm,
    xrightmargin=0.25cm,
    showstringspaces=false,
    showtabs=false,                  
    tabsize=2
}
\lstset{
  style=mystyle
}




\title{\attackNames: Stealing Intel Secrets from \\ SGX Enclaves via Speculative Execution}

\author{
{\rm Guoxing Chen, Sanchuan Chen, Yuan Xiao, Yinqian Zhang, Zhiqiang Lin, Ten H. Lai }\\
Department of Computer Science and Engineering\\
The Ohio State University\\
\{chen.4329, chen.4825, xiao.465\}@osu.edu\\ 
\{yinqian, zlin, lai\}@cse.ohio-state.edu
}




\maketitle

\subsection*{Abstract}

%
This paper presents \attackNames that exploit the recently disclosed CPU bugs to subvert the confidentiality and integrity  
of SGX enclaves. Particularly,  we show that when branch prediction of the enclave code can be influenced by programs outside the enclave, the control flow of the enclave program can be temporarily altered to execute instructions that lead to observable cache-state changes. An adversary observing such changes can learn secrets inside the enclave memory or its internal registers, thus completely defeating the confidentiality guarantee offered by SGX. To demonstrate the practicality of our \attackNames, we have systematically explored the possible attack vectors of branch target injection, approaches to win the race condition during enclave's speculative execution, and techniques to automatically search for code patterns required for launching the attacks. Our study suggests that \textit{any} enclave program could be vulnerable to \attackNames since the desired code patterns are available in most SGX runtimes (\eg, Intel SGX SDK, Rust-SGX, and Graphene-SGX). Most importantly, we have applied \attackNames to steal seal keys and attestation keys from Intel signed quoting enclaves. The seal key can be used to decrypt sealed storage outside the enclaves and forge valid sealed data; the attestation key can be used to forge attestation signatures. For these reasons, \attackNames practically defeat SGX's security protection. This paper also systematically evaluates Intel's existing countermeasures against \attackNames and discusses the security implications. 
\looseness=-1
%
 

\section{Introduction}

Software Guard eXtensions (SGX) is a set of micro-architectural extensions available in recent Intel processors. It is designed to improve the application security by removing the privileged code from the trusted computing base (TCB). At a high level, SGX provides software applications shielded execution environments, called \textit{enclaves}, to run private code and operate sensitive data, where both the code and data are isolated from the rest of the software systems. Even privileged software such as the operating systems and hypervisors are not allowed to directly inspect or manipulate the memory inside the enclaves.  Software applications adopting Intel SGX are partitioned into sensitive and non-sensitive components.  The sensitive components run inside the SGX enclaves (hence called \textit{enclave programs}) to harness the SGX protection, while non-sensitive components run outside the enclaves and interact with the system software. In additional to memory isolation, SGX also provides hardware-assisted memory encryption, remote attestation, and cryptographically sealed storage to offer comprehensive security guarantees.

Although SGX is still in its infancy, the promise of shielded execution has encouraged researchers and practitioners to develop various new applications to utilize these features (\eg,~\cite{Anati:2013:sgxseal, McKeen:2013:sgxisolate, Hoekstra:2013:sgxsolution, Schuster:2015:vc3,Zhang:2016:towncrier, Tramer:2016:SGP, Ohrimenko:2016:OMP, Tamrakar:2017:CGS,Zheng:2017:opaque}), and new software tools or frameworks (\eg,~\cite{Baumann:2015:haven, Arnautov:2016:scone, Strackx:2016:ariadne, Hunt:2016:ryoan, Tsai:2017:graphene, Kuvaiskii:2017:SMS,Weiser:2017:sgxio,Tychalas:2017:sgxcrypter,Shinde:2017:Panoply,Seo:2017:sgxshield, Matetic:2017:ROTE}) to help developers adopt this emerging programming paradigm. Most recently, SGX has been adopted by commercial public clouds, such as Azure confidential computing~\cite{AzureSGX}, aiming to protect cloud data security even with compromised operating systems or hypervisors, or even ``malicious insiders with administrative privilege''.

In SGX, the CPU itself, as part of the TCB, plays a crucial role in the security promises. However, the recently disclosed CPU vulnerabilities due to the out-of-order and speculative execution~\cite{googleprojectzero} have raised many questions and concerns about the security of SGX. Particularly, the so-called Meltdown~\cite{meltdown} and Spectre attacks~\cite{spectre} have demonstrated that an unprivileged application may exploit these vulnerabilities to extract memory content that is only accessible to privileged software. 
The developers have been wondering whether SGX will hold its original security promises after the disclosure of these hardware bugs~\cite{intelforum}. It is therefore imperative to answer this important question and understand its implications to SGX. 

As such, we set off our study with the goal of comprehensively understanding the security impact of these CPU vulnerabilities on SGX. Our study leads to the \attackNames, a new breed of the Spectre attacks on SGX. At a high level, \attack exploits the race condition between the injected, speculatively executed memory references, which lead to side-channel observable cache traces, and the latency of the branch resolution. We coin a new name for our SGX version of the Spectre attacks not only for the convenience of our discussion, but also to highlight the important differences between them, including the threat model, the attack vectors, the techniques to win the race conditions, and the consequences of the attacks. We will detail these differences in later sections. 

\attackNames are a new type of SGX side-channel attacks. Although it has already been demonstrated that by observing execution traces of an enclave program left in the CPU caches~\cite{Schwarz:2017:MGE, Brasser:2017:SGE, hahnel:2017:HRS, Gotzfried:2017:CAI}, branch target buffers~\cite{Lee:2017:IFC}, DRAM's row buffer contention~\cite{Wang:2017:LCD}, page-table entries~\cite{Van:2017:TYS, Wang:2017:LCD}, and page-fault exception handlers~\cite{Xu:2015:CAD, Shinde:2015:PYF}, a side-channel adversary with system privileges may \textit{infer} sensitive data from the enclaves, these traditional side-channel attacks are only feasible if the enclave program already has secret-dependent memory access patterns. In contrast, the consequences of \attackNames are far more concerning. 

\bheading{Our findings.} We show that \attackNames completely compromise the confidentiality and integrity of SGX enclaves. In particular, because vulnerable code patterns exist in most SGX runtime libraries (\eg, Intel SGX SDK, Rust-SGX, Graphene-SGX) and are difficult to  be eliminated, the adversary could perform \attackNames against \textit{any} enclave programs. We demonstrate end-to-end attacks to show that the adversary could learn the content of the enclave memory as well as its register values from a victim enclave developed by enclave developers (\ie, independent software vendors or ISVs). 

A even more alarming consequence is that \attackNames can be leveraged to steal secrets belonging to Intel SGX platforms, such as \textit{provisioning keys}, \textit{seal keys}, and \textit{attestation keys}. For example, we have demonstrated in an example that the adversary is able to extract the \textit{seal keys} of an enclave (or all enclaves belonging to the same ISV) when the key is being used. With the extracted seal key, our experiments suggest the enclave's sealed storage can be decrypted outside the enclave or even on a different machine; it can be further modified and re-encrypted to deceive the enclave, breaking both the confidentiality and integrity guarantees. 

Besides enclaves developed by ISVs, Intel's privately signed enclaves (\eg, the provisioning enclave and quoting enclave) are also vulnerable to \attackNames. As \attackNames have empowered a malicious OS to arbitrarily read enclave memory at any given time, any secrets provisioned by Intel's provisioning service (\eg, the attestation key) can be leaked as long as they temporarily appear in the enclave memory. We have demonstrated that \attackNames are able to read memory from the quoting enclave developed by Intel and extract Intel's seal key, which can be used to decrypt the sealed EPID blob to extract the attestation key (\ie, EPID private key). 

\bheading{Security implications.}
Intel's solutions to \attackNames are twofold: First, Intel has released a microcode update (\ie, indirect branch restricted speculation, or IBRS) to prevent branch injection attacks. Our experiments shows that IBRS could cleanse the branch prediction history at the enclave boundary, thus rendering our \attackNames ineffective. Second, Intel's remote attestation service, which arbitrates every attestation request from the ISV, responses to the attestation signatures generated from unpatched CPUs with an error message indicating outdated CPU security version number 
(\texttt{CPUSVN}). 

Nevertheless, the security implications are as follows: First, any secret allowed to be provisioned to an unpatched processor can be leaked, which includes secrets provisioned before the microcode update and secrets provisioned without attestation. Second, the EPID private key used for remote attestation can be extracted by the attacker, which allows the attacker to emulate an enclave environment entirely outside the enclave while providing a valid (though outdated) signature.

\bheading{Responsible disclosure.} We have disclosed our study to the security team at Intel before releasing our study to the public. The tool for scanning vulnerabilities in enclave code has been open sourced.


\bheading{Contributions.} This paper makes the following contributions.

\begin{packeditemize}

\item \textit{Systematic studies of a timely issue.} We provide the first comprehensive exploration of the impacts of the recent micro-architect-ural vulnerabilities on the security of SGX. 

\item \textit{New techniques to enable SGX attacks.} We develop several new techniques that enable attacks against any enclave programs, including symbolic execution of SDK runtime binaries for vulnerability detection and combination of various side-channel techniques for winning the race conditions.  

\item \textit{The first attack against Intel signed enclaves.} To the best of our knowledge, the attacks described in this paper are the first to extract Intel secrets (\ie, attestation keys) from Intel signed quoting enclaves.

\item\textit{Security implications for SGX.} Our study concludes that SGX processors with these hardware vulnerabilities are no longer trustworthy, urging the enclave developers to add vulnerability verification into their development.

 
\end{packeditemize}

%
%
\bheading{Roadmap.} \secref{sec:background} introduces key concepts of Intel processor micro-architectures to set the stage of our discussion. \secref{sec:threat} discusses the threat model. \secref{sec:attack} presents a systematic exploration of attack vectors in enclaves and techniques that enable practical attacks. \secref{sec:gadget} presents a symbolic execution tool for automatically searching instruction gadgets in enclave programs. \secref{sec:exploit} shows end-to-end \attackNames against enclave runtimes that lead to a complete breach of enclave confidentiality. \secref{sec:counter} discusses and evaluates countermeasures against the attacks. \secref{sec:related} discusses related work and \secref{sec:conclude} concludes the paper.

\section{Background}
\label{sec:background}

\subsection{Intel Processor Internals}

\bheading{Out-of-order execution.} 
Modern CPUs implement deep pipelines, so that multiple instructions can be executed at the same time. Because instructions do not take equal time to complete, the order of the instructions' execution and their order in the program may differ. This form of out-of-order execution requires taking special care of instructions whose operands have inter-dependencies, as these instructions may access memory in orders constrained by the program logic. To handle the potential data hazards, instructions are retired in order, resolving any inaccuracy due to the out-of-order execution at the time of retirement.

\bheading{Speculative execution.}
Speculative execution shares the same goal as out-of-order execution, but differs in that speculation is made to speed up the program's execution when the control flow or data dependency of the future execution is uncertain. One of the most important examples of speculative execution is branch prediction. When a conditional or indirect branch instruction is met, because checking the branch condition or resolving branch targets may take time, predictions are made, based on its history, to prefetch instructions first. If the prediction is true, speculatively executed instructions may retire; otherwise mis-predicted execution will be re-winded. The micro-architectural component that enables speculative execution is the branch prediction unit (BPU), which consists of several hardware components that help predict conditional branches, indirect jumps and calls, and function returns. For example, branch target buffers (BTB) are typically used to predict indirect jumps and calls, and return stack buffers (RSB) are used to predict near returns. These micro-architectural components, however, are shared between software running on different security domains (\eg, user space vs. kernel space, enclave mode vs. non-enclave mode), thus leading to the security issues that we present in this paper. \looseness=-1

\bheading{Implicit caching.}
Implicit caching refers to the caching of memory elements, either data or instructions, that are not due to direct instruction fetching or data accessing. Implicit caching may be caused in modern processors by ``aggressive prefetching, branch prediction, and TLB miss handling''~\cite{IntelDevelopmentManual}. For example, mis-predicted branches will lead to the fetching and execution of instructions, as well as data memory reads or writes from these instructions, that are not intended by the program. Implicit caching is one of the root causes of the CPU vulnerabilities studied in this paper.

\subsection{Intel SGX}

Intel SGX is a hardware extension in recent Intel processors aiming to offer stronger application security by providing primitives such as memory isolation, memory encryption, sealed storage, and remote attestation. An important concept in SGX is the secure enclave. 
An enclave is an execution environment created and maintained by the processor so that only applications running in it have a dedicated memory region that is protected from all other software components. Both confidentiality and integrity of the memory inside enclaves are protected from the untrusted system software.

\bheading{Entering and exiting enclaves.}
To enter the enclave mode, the software executes the \texttt{EENTER} leaf function by specifying the address of Thread Control Structure (TCS) inside the enclave. TCS holds the location of the first instruction to execute inside the enclave. Multiple TCSs can be defined to support multi-threading inside the same enclave. Registers used by the untrusted program may be preserved after \texttt{EENTER}. The enclave runtime needs to determine the proper control flow depending on the register values (\eg, differentiating \texttt{ECall} from \texttt{ORet}). 
  \looseness=-1

\bheading{Asynchronous Enclave eXit (AEX).}
When interrupts, exceptions, and VM exits happen during the enclave mode, the processor will save the execution state in the State Save Area (SSA) of the current enclave thread, and replace it with a synthetic state to prevent information leakage. After the interrupts or exceptions are handled, the execution will be returned (through \texttt{IRET}) from the kernel to an address external to enclaves, which is known as Asynchronous Exit Pointer (AEP). The \texttt{ERESUME} leaf function will be executed to transfer control back to the enclave by filling the \texttt{RIP} with the copy saved in the SSA.


\bheading{CPU security version.}
Intel SGX uses a CPU Security Version Number (\texttt{CPUSVN)} to reflect the processor's microcode update version, and considers all SGX implementations with older \texttt{CPUSVN} to be untrustworthy. Whenever security vulnerabilities are fixed with microcode patches, the \texttt{CPUSVN} will be updated.

\bheading{Sealed storage.}
Enclaves can encrypt and integrity-protect some secrets via a process, called \textit{sealing}, to store the secrets outside the enclave, \eg, on a non-volatile memory. The encryption key used during the sealing process, is called the \textit{seal key}, which is derived via \texttt{EGETKEY} instruction. A \texttt{CPUSVN} has to be specified when deriving a seal key. While it is allowed to derived seal keys with \texttt{CPUSVN}s older than current \texttt{CPUSVN} to access legacy sealed secrets, deriving seal keys with newer \texttt{CPUSVN} is forbidden to prevent attack such as rolling back the microcode to a vulnerable version to steal the secrets sealed with newer \texttt{CPUSVN}.



\bheading{Remote Attestation.}
SGX remote attestation is used by enclaves to prove to the ISV (\ie, the enclave developer) that a claimed enclave is running inside an SGX enabled processor. An anonymous signature scheme, called Intel \textit{Enhanced Privacy ID} (EPID), is used to produce the attestation signature, which could be verified later by the Intel attestation service.
The attestation key (\ie, EPID private key) cannot be directly accessed by an attested enclave, otherwise a malicious enclave could generate any valid attestation signature to deceive the remote party. Hence, Intel issues two privileged enclaves, called the \textit{Provisioning Enclave} (PvE) and the \textit{Quoting Enclave} (QE) to manage the attestation key and sign attestation data. Specifically, the provisioning enclave communicates with Intel provisioning service to obtain an attestation key and seals it on a non-volatile memory; the quoting enclave could unseal the sealed attestation key and produce attestation signature on behalf of an attested enclave. Note that during the attestation process, Intel attestation service could also verify the \texttt{CPUSVN} of the SGX platform running the attested enclave, and notify the ISV if the \texttt{CPUSVN} is outdated.

\subsection{Cache Side Channels}

Cache side channels leverage the timing difference between cache hits and cache misses to infer the victim's memory access patterns. Typical examples of cache side-channel attacks are \primeprobe and \flushreload attacks. In \primeprobe attacks~\cite{Percival:2005:CMF, Osvik:2006:CAC, Zhang:2012:CSC,Neve:2006:AAC,Aciicmez:2007:YMA,
Tromer:2010:ECA, Liu:2015:LCS, irazoqui:2015:shared}, by pre-loading cache lines in a cache set, the adversary expects that her future memory accesses (to the same memory) will be served by the cache, unless evicted by the victim program. Therefore, cache misses will reveal the victim's cache usage of the target cache set. In \flushreload attacks~\cite{Gullasch:2011:CGB, Yarom:2014:FRH, Yarom:2014:ROE, Benger:2014:JLB, Zhang:2014:CSA,Irazoqui:2014:WMF}, the adversary shares some physical memory pages (\eg, through dynamic shared libraries) with the victim. By issuing \texttt{clflush} on certain virtual
address that are mapped to the shared pages, the adversary can flush the shared cache lines out of the entire
cache hierarchy. Therefore, \Reload{s} of these cache lines will be slower because of cache misses, unless they have been loaded by the victim into the cache. In these ways, the victim's memory access patterns can be revealed to the adversary.

\section{Threat Model}
\label{sec:threat}

In this paper, we consider an adversary with the system privilege of the machine that runs on the processor with SGX support. Specifically, we assume the adversary has the following capabilities. 

\begin{packeditemize}


\item \textit{Complete OS Control:} We assume the adversary has complete control of the entire OS, including re-compiling of the OS kernel and rebooting of the OS with arbitrary argument as needed. 

\item \textit{Interacting with the targeted enclave:} We assume the adversary is able to launch the targeted enclave program with a software program under her control. This means the arguments of \texttt{ECall}s and return values of \texttt{OCall}s are both controlled by the adversary. 

\item \textit{Launching and controlling another enclave:} we assume the adversary is able to run another enclave that she completely controls in the same process or another process. This implies that the enclave can poison any BTB entries used by the targeted enclave. 

\end{packeditemize} 


The goal of the attack is to learn the memory content inside the enclave. We assume the binary code of the targeted enclave program is already known to the adversary and does not change during the execution. Therefore, we assume that the adversary is primarily interested in learning the secret data inside the enclaves after the enclave has been initialized (\eg, generating secrets from random values or downloading secrets from the enclave owners.) 

\begin{figure}[t]
    \centering
	\includegraphics[width=0.45\textwidth]{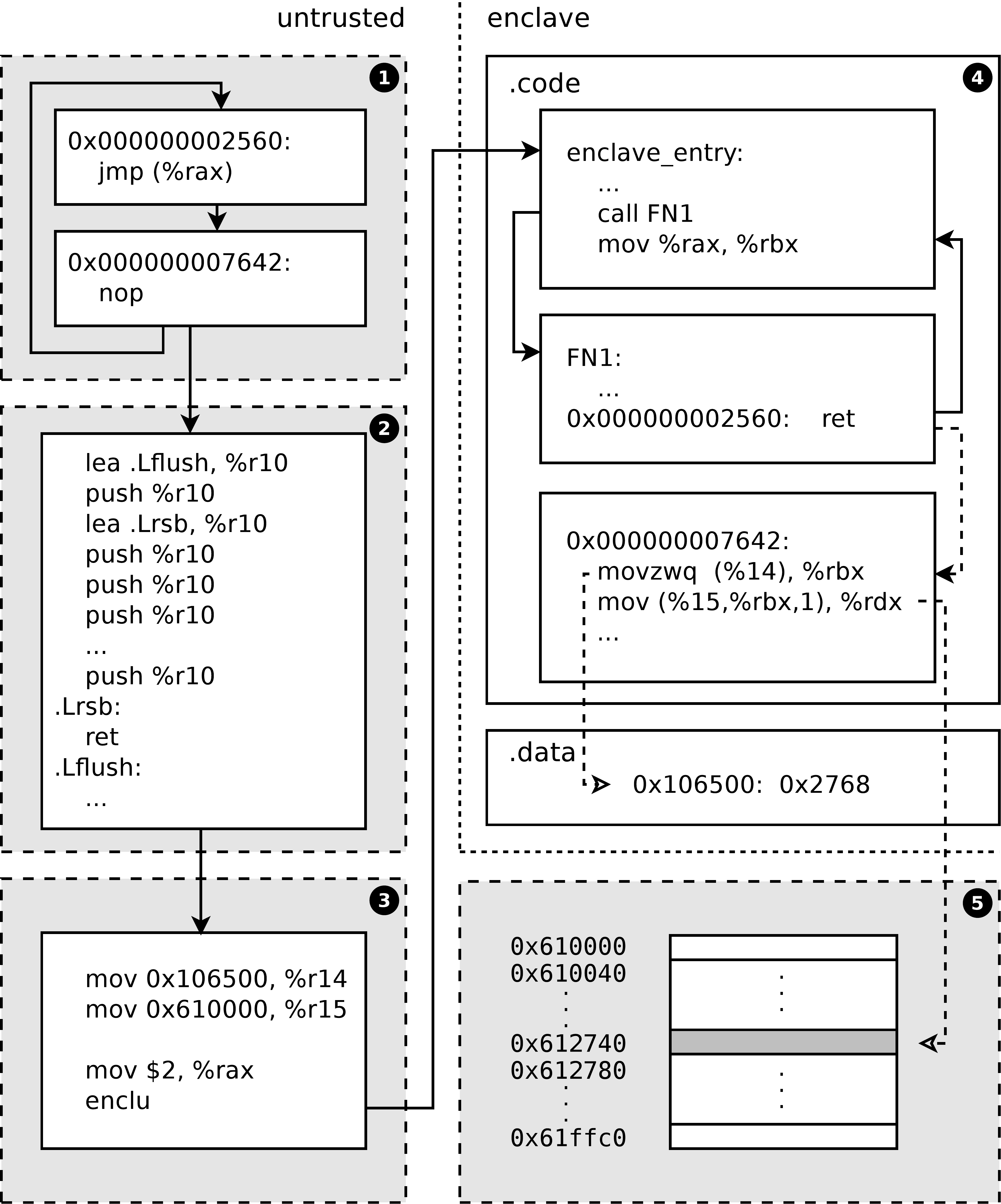}     \vspace{-0.1in}
	\caption{A simple example of \attackNames. The gray blocks represent code or data outside the enclave. The white blocks represent enclave code or data.}
\vspace{-0.1in}
	\label{fig:example}
\end{figure}


\section{\attackNames}
\label{sec:attack}


\subsection{A Simple Example}

The basic idea of an \attackName is illustrated in \autoref{fig:example}. There are 5 steps in an \attackName: 

\st{182} is to poison the branch target buffer, such that when the enclave program executes a branch instruction at a specific address, the predicted branch target is the address of enclave instructions that may leak secrets.
For example, in \figref{fig:example}, to trick the \texttt{ret} instruction at address $0$x$02560$ in the enclave to speculatively return to the secret-leaking instructions located at address $0$x$07642$, the code to poison the branch prediction executes an indirect jump from the source address $0$x$02560$ to the target address $0$x$07642$ multiple times. We will discuss branch target injection in more details 
in \secref{ss:injection}.  \looseness=-1

\st{183} is to prepare for a CPU environment to increase the chance of speculatively executing the secret-leaking instructions before the processor detects the mis-prediction and flushes the pipeline. Such preparation includes flushing the victim's branch target address (to delay the retirement of the targeted branch instruction or return instruction) and depleting the RSB (to force the CPU to predict return address using the BTB).  Flushing branch targets cannot use the \texttt{clflush} instruction, as the enclave memory is not accessible from outside (We will discuss alternative approaches in \secref{ss:race}). The code for depleting the RSB (shown in \figref{fig:example}) pushes the address of a \texttt{ret} instructions 16 times and returns to itself repeatedly to drain all RSB entries. 

\st{184} is to set the values of registers used by the speculatively executed secret-leaking instructions, such that they will read enclave memory targeted by the adversary and leave cache traces that the adversary could monitor. In this simple example, the adversary sets \texttt{r14} to $0$x$106500$,  the address of a $2$-byte secret inside the enclave, and sets \texttt{r15} to $0$x$610000$, the base address of a monitored array outside the enclave. The \texttt{enclu} instruction with \texttt{rax}=2 is executed to enter the enclave. We will discuss methods to pass values into the enclaves in \secref{ss:registers}. \looseness=-1

\st{185} is to actually run the enclave code. Because of the BTB poisoning, instructions at address $0$x$07642$ will be executed speculatively when the target of the \texttt{ret} instruction at address $0$x$02560$ is being resolved. The instruction ``\texttt{movzwq (\%r14), \%rbx}'' loads the $2$-byte secret data into \texttt{rbx}, and ``\texttt{mov (\%r15, \%rbx, 1), \%rdx}'' touches one entry of the monitored array dictated by the value of \texttt{rbx}.  \looseness=-1

\st{186} is to examine the monitored array using a \flushreload side channel and extract the secret values. Techniques to do so are discussed in details in \secref{ss:leaking}.





\subsection{Injecting Branch Targets into Enclaves}
\label{ss:injection}



The branch prediction units in modern processors typically consists of:
 \looseness=-1
 
\begin{packeditemize}

\item \textit{Branch target buffer:}
When an indirect jump/call or a conditional jump is executed, the target address will be cached in the BTB. The next time the same indirect jump/call is executed, the target address in the BTB will be fetched for speculative execution. Modern x$86$-$64$ architectures typically support $48$-bit virtual address and $40$-bit physical address~\cite{IntelDevelopmentManual, Keltcher:2003:AOP}. For space efficiency, many Intel processors, such as Skylake, uses the lower $32$-bit of a virtual address as the index and tag of a BTB entry. 
\item \textit{Return stack buffer:} 
When a near \texttt{Call} instruction with non-zero displacement\footnote{\texttt{Call} instructions with zero displacement will not affect the RSB, because they are common code constructions for obtaining the current \texttt{RIP} value. 
These zero displacement calls do not have matching returns.} is executed, an entry with the address of the instruction sequentially following it will be created in the return stack buffer (RSB). 
The RSB is not affected by far \texttt{Call}, far \texttt{Ret}, or \texttt{Iret} instructions. Most processors that implement RSB have $16$ entries~\cite{fog2017microarchitecture}. On Intel Skylake or later processors, when RSB underflows, BTBs will be used for prediction instead.
 \end{packeditemize}

\begin{figure}[t]
  	\centering
   	\includegraphics[width=0.45\textwidth]{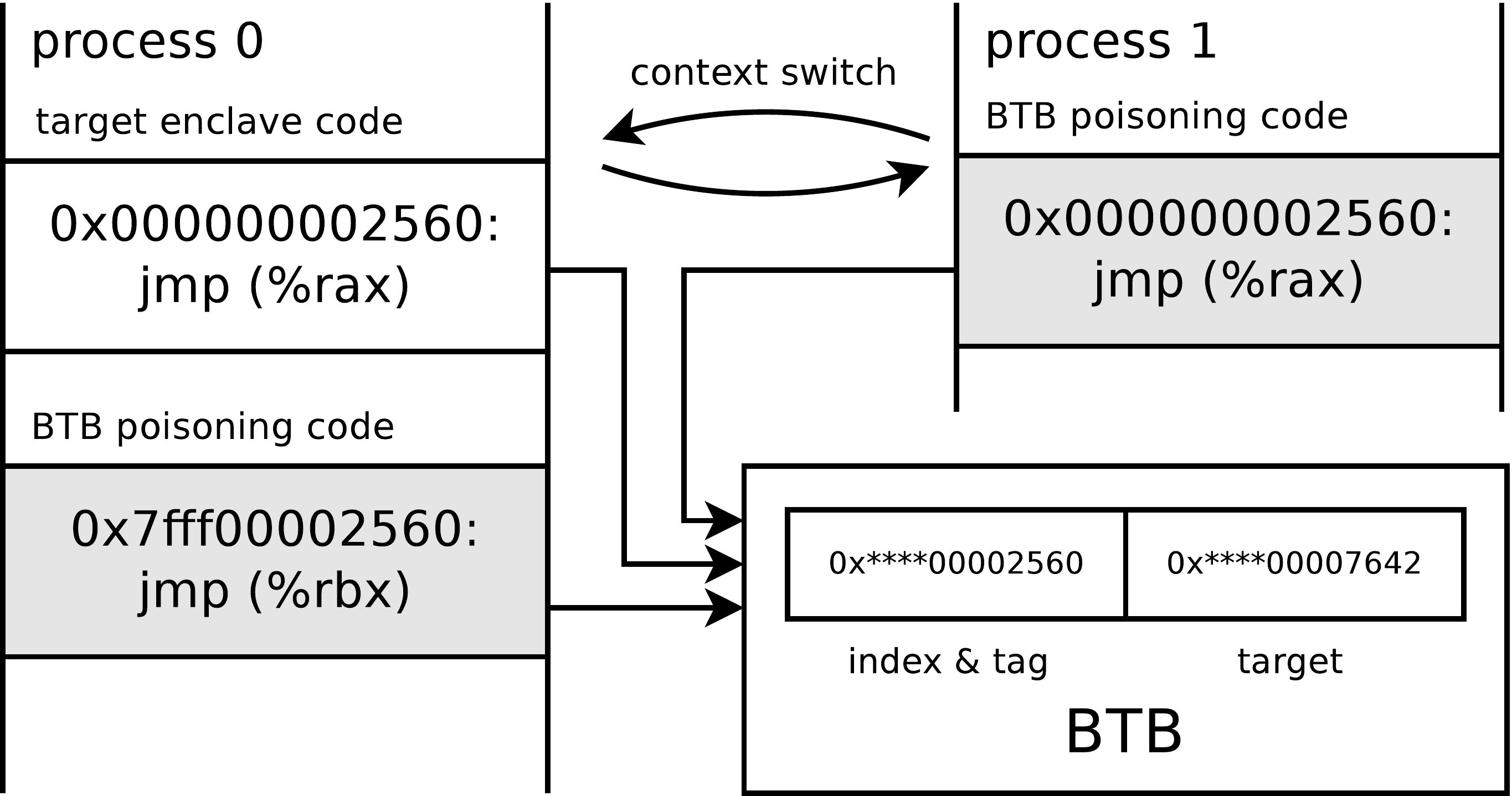}
    \vspace{-0.1in}
  	\caption{Poisoning BTB from the Same Process or A Different Process}
	\label{fig:btb}
\vspace{-0.1in}
\end{figure}

\bheading{Poisoning BTBs from outside.}
To temporarily alter the control-flow of the enclave code by injecting branch targets, the adversary needs to run BTB poisoning code outside the targeted enclave, which could be done in one of the following ways (as illustrated in \figref{fig:btb}). 


\begin{packeditemize}
\item \textit{Branch target injection from the same process.} 
 The adversary could poison the BTB by using code outside the enclave but in the same process. Since the BTB uses only the lower 32 bits of the source address as BTB indices and tags, the adversary could reserve a $2^{32} = 4$GB memory buffer, and execute an indirect jump instruction (within the buffer) whose source address (\ie, $0$x$7$fff$00002560$) is the same as the branch instruction in the target enclave (\ie, $0$x$02560$) in the lower $32$ bits, and target address (\ie, $0$x$7$fff$00007642$) is the same as the secret-leaking instructions (\ie, $0$x$07642$) inside the target enclave in the lower $32$ bits.  \looseness=-1

\item \textit{Branch target injection from a different process.} 
The adversary could inject the branch targets from a different process. Although this attack method requires a context switch in between of the execution of the BTB poisoning code and targeted enclave program, the advantage of this method is that the adversary could encapsulate the BTB poisoning coding into another enclave that is under his control. This allows the adversary to perfectly shadow the branch instructions of the targeted enclave program (\ie, matching all bits in the virtual addresses). 
\end{packeditemize}

It is worth noting that address space layout randomization can be disabled by adversary to facilitate the BTB poisoning attacks. On a Lenovo Thinkpad X1 Carbon ($4$th Gen) laptop with an Intel Core i$5$-$6200$U processor (Skylake), we have verified that for indirect jump/call, the BTB could be poisoned either from the same process, or a different process. For the return instructions, we only observed successful poisoning using a different process (\ie, perfect branch target matching). 
To force return instructions to use BTB, the RSB needs to be depleted before executing the target enclave code. Interestingly, as shown in \figref{fig:example}, a near call is made in \texttt{enclave\_entry}, which could have filled the RSB, but we still could inject the return target of the return instruction at $0$x$02560$ with BTB. We speculate that this is a architecture-specific implementation. A more reliable way to deplete the RSB is through the use of AEX as described in \secref{sec:exploit:ssa}.

\looseness=-1

\looseness=-1

\subsection{Controlling Registers in Enclaves}
\label{ss:registers}
Because all registers are restored by hardware after \texttt{ERESUME}, the adversary is not able to control any registers inside the enclave when the control returns back to the enclave after an AEX. In contrast, most registers can be set before the \texttt{EENTER} leaf function and remain controlled by the adversary after entering the enclave mode until modified by the enclave code. Therefore, the adversary might have a chance to control some registers in the enclave after an \texttt{EENTER}. 

\begin{figure}
\centering
\includegraphics[width=0.9\columnwidth]{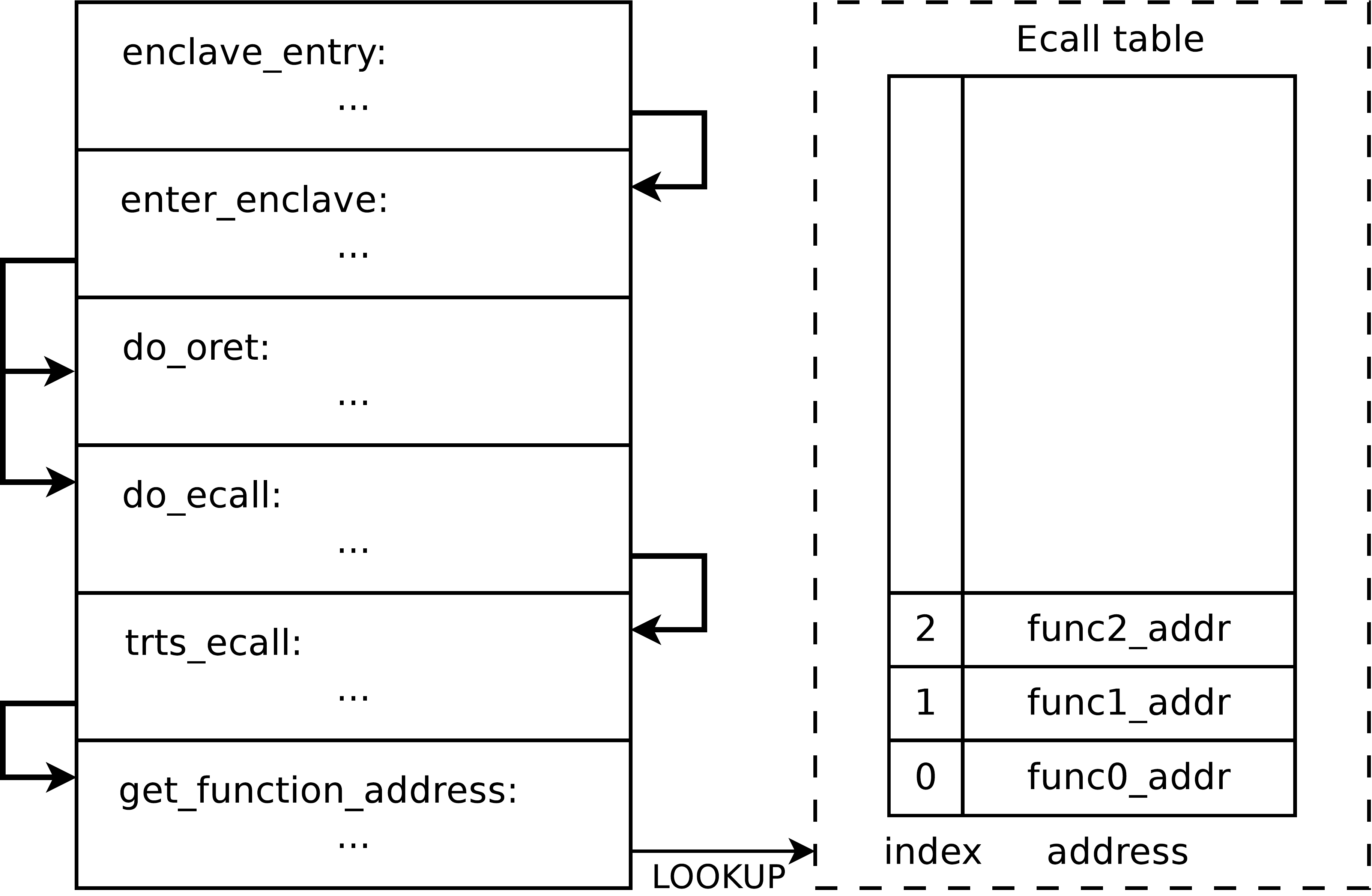}
\vspace{-0.1in}
\caption{EENTER and ECall Table Lookup}
\label{fig:ecall}
\vspace{-0.1in}
\end{figure}

The SGX developer guide~\cite{SGXprogramming} defines \texttt{ECall} and \texttt{OCall} to specify the interaction between the enclave and external software. An \texttt{ECall}, or ``Enclave Call'', is a function call to enter enclave mode; an \texttt{OCall}, or ``Outside Call'', is a function call to exit the enclave mode. Returning from an \texttt{OCall} is called an \texttt{ORet}. Both \texttt{ECall}s and \texttt{ORet}s are implemented through \texttt{EENTER} by the SGX SDK. As shown in \figref{fig:ecall}, the function \texttt{enter\_enclave} is called by the enclave entry point, \texttt{enclave\_entry}. Then depending on the value of the \texttt{edi} register, \texttt{do\_ecall} or \texttt{do\_oret} will be called. The \texttt{do\_ecall} function is triggered to call \texttt{trts\_ecall} and \texttt{get\_function\_address} in a sequence and eventually look up the \texttt{Ecall} table. Both \texttt{Ecall} and \texttt{ORet} can be exploited to control registers in enclaves.


\subsection{Leaking Secrets via Side Channels}
\label{ss:leaking}
The key to the success of \attackNames lies in the artifact that speculatively executed instructions trigger implicit caching, which is not properly rewinded when these incorrectly issued instructions are discarded by the processor. Therefore, these side effects of speculative execution on the CPU caches can be leveraged to leak information from inside the enclave. 

Cache side-channel attacks against enclave programs have been studied recently~\cite{Schwarz:2017:MGE, Brasser:2017:SGE, hahnel:2017:HRS, Gotzfried:2017:CAI}, all of which demonstrated that a program runs outside the enclave may use \primeprobe techniques~\cite{Tromer:2010:ECA} to extract secrets from the enclave code, only if the enclave code has secret-dependent memory access patterns. Though more fine-grained and less noisy, \flushreload techniques~\cite{Yarom:2014:FRH} cannot be used in SGX attacks because enclaves do not share memory with the external world.

Different from these studies, however, \attackNames may leverage these less noisy \flushreload side channels to leak information. Because the enclave code can access data outside the enclave directly, an \attackName may force the speculatively executed memory references inside enclaves to touch memory location outside the enclave, as shown in \autoref{fig:example}. The adversary can flush an array of memory before the attack, such as the array from address $0$x$610000$ to $0$x$61$ffff, and then reload each entry and measure the reload time to determine if the entry has been touched by the enclave code during the speculative execution.

Other than cache side-channel attacks, previous work has demonstrated BTB side-channel attacks, TLB side-channel attacks, DRAM-cache side-channel attacks, and page-fault attacks against enclaves. In theory, some of these venues may also be leveraged by \attackNames. For instance, although TLB entries used by the enclave code will be flushed when exiting the enclave mode, a \primeprobe-based TLB attack may learn that a TLB entry has been created in a particular TLB set when the program runs in the enclave mode. Similarly, BTB and DRAM-cache side-channel attacks may also be exploitable in this scenario. However, page-fault side channels cannot be used in \attackNames because the speculatively executed instructions will not raise exceptions.

\subsection{Winning a Race Condition}
\label{ss:race}

\begin{figure}
\centering
\includegraphics[width=0.9\columnwidth]{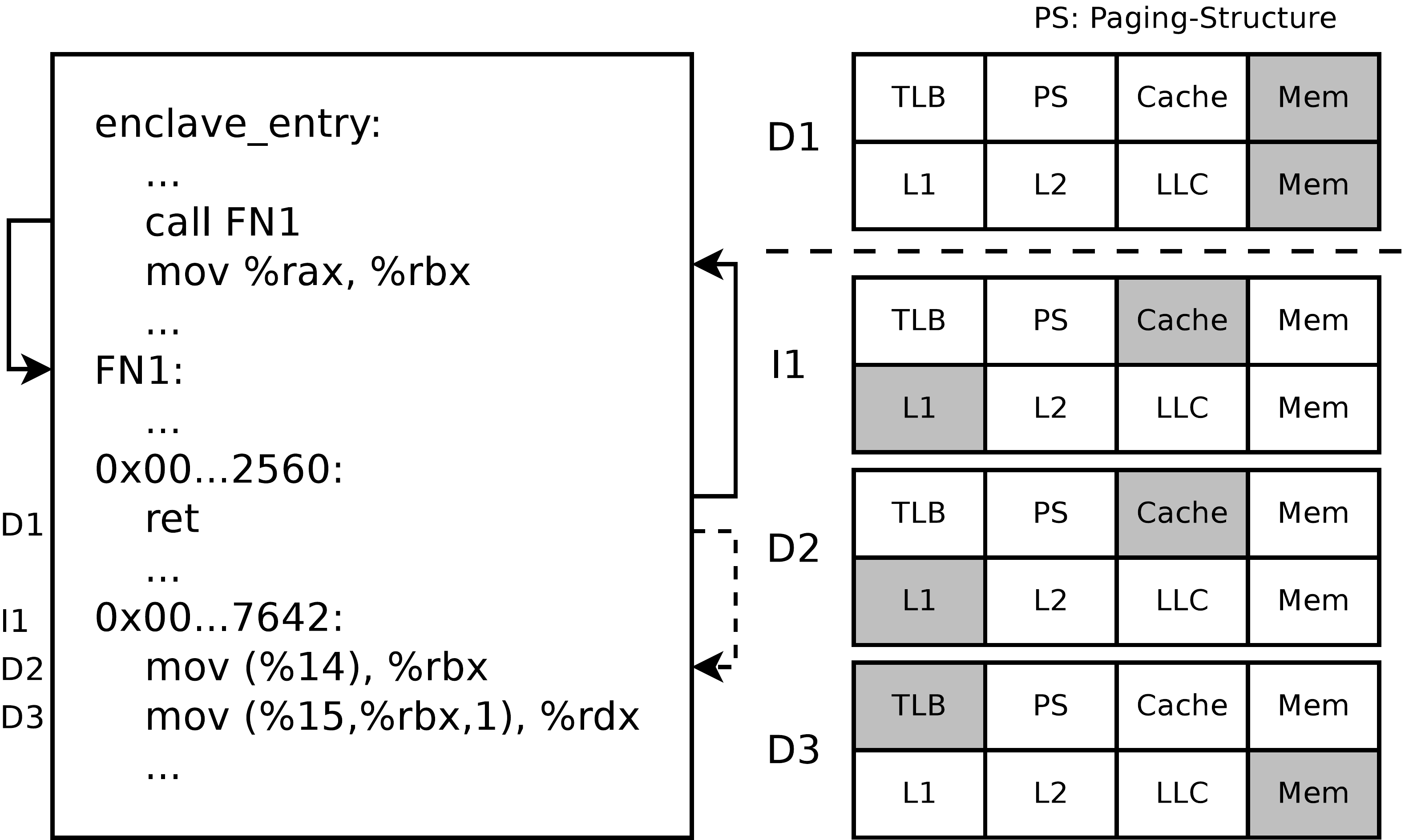}
\vspace{-0.1in}
\caption{Best scenarios for wining a race condition. Memory accesses \texttt{D1}, \texttt{I1}, \texttt{D2}, \texttt{D3} are labeled next to the related instructions. The address translation and data accesses are illustrated on the right: The $4$ blocks on top denote the units holding the address translation information, including TLBs, paging structures, caches (for PTEs), and the memory; the $4$ blocks at the bottom denote the units holding data/instruction. The shadow blocks represent the units from which the address translation or data/instruction access are served.}
\label{fig:race}
\vspace{-10pt}
\end{figure}

At the core of an \attackName is a race between the execution of the branch instruction and the speculative execution: data leakage will only happen when the branch instruction retires later than the speculative execution of the secret-leaking code. \figref{fig:race} shows a desired scenario for wining such a race condition in an \attackName: The branch instruction has one data access \texttt{D1}, while the speculative execution of the secret-leaking code has one instruction fetch \texttt{I1} and two data accesses \texttt{D2} and \texttt{D3}. To win the race condition, the adversary should ensure that the memory accesses of \texttt{I1}, \texttt{D2} and \texttt{D3} are fast enough. However, because \texttt{I1} and \texttt{D2} fetch memory inside the enclave, and as TLBs and paging structures used inside the enclaves are flushed at AEX or \texttt{EEXIT}, the adversary could at best perform the address translation of the corresponding pages from caches (\ie, use cached copies of the page table). Fortunately, it can be achieved by performing the attack \st{185} in \figref{fig:example} multiple times. It is also possible to preload the instructions and data used in \texttt{I1} and \texttt{D2} into the L1 cache to further speed up the speculative execution. As \texttt{D3} accesses memory outside the enclave, it is possible to preload the TLB entry of the corresponding page. However, data of \texttt{D3} must be loaded from the memory.

Meanwhile, the adversary should slow down \texttt{D1} by forcing its address translation and data fetch to happen in the memory. 
However, this step has been proven technically challenging. First, it is difficult to effectively flush the branch target (and the address translation data) to memory without using \texttt{clflush} instruction. Second, because the return address is stored in the stack frames, which is very frequently used during the execution, evicting return addresses must be done frequently. In the attack described in \secref{sec:exploit}, we leveraged an additional page fault to suspend the enclave execution right before the branch instruction and flush the return target by evicting all cache lines in the same cache set. 


\ignore{
\subsection{Proof-of-Concept Examples}
We next demonstrate a few examples of \attackNames in SGX enclaves to prove the concept. Particularly, we demonstrate (1) how the conditional branches can be mis-predicted to allow execution of alternative branches, (2) how indirect branches can be injected to alter the control flow of the enclave code to arbitrary address inside the enclave, and (3) how the return stack buffers can be poisoned inside SGX enclaves. It is worth noting that we intentionally designed these examples to target ``toy'' enclave applications, in order to demonstrate their exploitability. We will present methods to scan enclave code to identify both type I and type II gadgets in various SGX runtimes in \secref{sec:scan} and demonstrate some of these attacks against real-world enclave code in \secref{sec:exploit}.
\subsubsection{Conditional Branch Injection}
\yz{Guoxing, put your dummy example here, discuss methods to exploit, and success rate.}
\subsubsection{Indirect Branch Target Injection}
\yz{Guoxing, put your dummy examples here, both indirect jmp and call, discuss methods to exploit, and success rate.}
\subsubsection{Return Address Injection}
\yz{Guoxing, show that rsb can be exploited if the return address points to an address inside the enclave; show negative results when the return address is outside the enclave.}
}

\section{Attack Gadgets Identification}
\label{sec:gadget}



In this section, we show that any enclave programs developed with existing SGX SDKs are vulnerable to \attackNames. In particular, we have developed an automated program analysis tool 
that symbolically executes the enclave code to examine code patterns in the SGX runtimes, and have identified those code patterns in every runtime library we have examined, including Intel's SGX SDK~\cite{sgxsdk}, Graphene-SGX~\cite{Tsai:2017:graphene}, Rust-SGX~\cite{Ding:2017:rust-sgx}. In this section, we present how we search these gadgets in greater detail.

\subsection{Types of Gadgets}

In order to launch \attackNames, two types of code patterns are needed. The first type of code patterns consists of a branch instruction that can be influenced by the adversary and several registers that are under the adversary's control when the branch instruction is executed. The second type of code patterns consists of two memory references sequentially close to each other and collectively reveals some enclave memory content through cache side channels. Borrowing the term used in return-oriented programming~\cite{Shacham:2007:rop} and Spectre attacks~\cite{spectre}, we use \textit{gadgets} to refer to these patterns. More specifically, we name them \textit{Type-I gadgets} and \textit{Type-II gadgets}, respectively.

\subsubsection{Type-I gadgets: branch target injection}
\label{s:type1}

Unlike the typical ROP gadget, we consider a gadget to be just a sequence of instructions that are executed sequentially during one run of the enclave program and they may not always be consecutive in the memory layout. A Type-I gadget is such an instruction sequence that starts from the entry point of \texttt{EENTER} (dubbed \texttt{enclave\_entry}) and ends with one of the following instructions: (1) near indirect jump, (2) near indirect call, or (3) near return. \texttt{EENTER} is the only method for the adversary to take control of registers inside enclaves. During an \texttt{EENTER}, most registers are preserved by the hardware; 
they are left to be sanitized by the enclave software. If any of these registers are not overwritten by the software before one of the three types of branch instructions are met, a Type-I gadget is found. 

\begin{figure}[t]
\footnotesize
\begin{lstlisting}[frame=single, caption={An Example of a Type-I Gadget}, label={lst:gadgetI}]
0000000000003662 <enclave_entry>:
3662: cmp    $0x0,%rax
3666: jne    3709 <enclave_entry+0xa7>
366c: xor    %rdx,%rdx
366f: mov    %gs:0x8,%rax
3676:	00 00 
3678: cmp    $0x0,%rax
367c: jne    368d <enclave_entry+0x2b>
367e: mov    %rbx,%rax
3681: sub    $0x10000,%rax
3687: sub    $0x2b0,%rax
368d: xchg   %rax,%rsp
368f: push   %rcx
3690: push   %rbp
3691: mov    %rsp,%rbp
3694: sub    $0x30,%rsp
3698: mov    %rax,-0x8(%rbp)
369c: mov    %rdx,-0x18(%rbp)
36a0: mov    %rbx,-0x20(%rbp)
36a4: mov    %rsi,-0x28(%rbp)
36a8: mov    %rdi,-0x30(%rbp)
36ac: mov    %rdx,%rcx
36af: mov    %rbx,%rdx
36b2: callq  1f20 <enter_enclave>
...

0000000000001f20 <enter_enclave>:
1f20: push   %r13
1f22: push   %r12
1f24: mov    %rsi,%r13
1f27: push   %rbp
1f28: push   %rbx
1f29: mov    %rdx,%r12
1f2c: mov    %edi,%ebx
1f2e: mov    %ecx,%ebp
1f30: sub    $0x8,%rsp
1f34: callq  b60 <sgx_is_enclave_crashed>
...

0000000000000b60 <sgx_is_enclave_crashed>:
b60: sub    $0x8,%rsp
b64: callq  361b <get_enclave_state>
...

000000000000361b <get_enclave_state>:
361b: lea    0x213886(%rip),%rcx     # 216ea8 <g_enclave_state>
3622: xor    %rax,%rax
3625: mov    (%rcx),%eax
3627: retq   
\end{lstlisting}
\end{figure}

An example of a Type-I gadget is shown in \autoref{lst:gadgetI}, which is excerpted from \texttt{libsgx\_trts.a} of Intel SGX SDK. 
In particular, line $49$ in \autoref{lst:gadgetI} is the first return instruction encountered by an enclave program after \texttt{EENTER}. When this near return instruction is executed, several registers can still be controlled by the adversary, including \texttt{rbx}, \texttt{rdi}, \texttt{rsi}, \texttt{r8}, \texttt{r9}, \texttt{r10}, \texttt{r11}, \texttt{r14}, and \texttt{r15}.  

\bheading{Gadget exploitability.} The exploitability of a Type-I gadget is determined by the number of registers that are controlled (both directly or indirectly) by the adversary at the time of the execution of the branch instruction. The more registers that are under control of the adversary, the higher the exploitability of the gadget. Highly exploitable Type-I gadgets mean less restriction on the Type-II gadgets in the exploits.

\subsubsection{Type-II gadgets: secret extraction}
\label{s:type2}

A Type-II gadget is a sequence of instructions that starts from a memory reference instruction that loads data in the memory pointed to by register \texttt{regA} into register \texttt{regB}, and ends with another memory reference instruction whose target address is determined by the value of \texttt{regB}. When the control flow is redirected to a Type-II gadget, if \texttt{regA} is controlled by the adversary, the first memory reference instruction will load \texttt{regB} with the value of the enclave memory chosen by the adversary. Because the entire Type-II gadget is speculatively executed and eventually discarded when the branch instruction in the Type-I gadget retires, the secret value stored in \texttt{regB} will not be learned by the adversary directly. However, as the second memory reference will trigger the implicit caching, the adversary can use a \flushreload side channel to extract the value of \texttt{regB}.

\begin{figure}[t]
\begin{lstlisting}[frame=single, caption={An Example of a Type-II Gadget},label={lst:gadgetII}]
0000000000005c10 <dlfree>:
...
607f: mov    0x38(%rsi),%edi
6082: mov    %rdi,%rcx
6085: lea    (%rbx,%rdi,8),%rdi
6089: cmp    0x258(%rdi),%rsi
...
\end{lstlisting}
\end{figure}

An example of a Type-II gadget is illustrated in \autoref{lst:gadgetII}, which is excerpted from 
the \texttt{libsgx\_tstdc.a} library of Intel SGX SDK. 
Assuming \texttt{rsi} is a register controlled by the adversary, the first instruction (line 3) reads the content of memory address pointed to by \texttt{rsi+0x38} to \texttt{edi}. Then the value of \texttt{rbx}+\texttt{rdi}$\times 8$ is stored in \texttt{rdi} (line 5). Finally, the memory address at \texttt{rdi+0x258} is loaded to be compared with \texttt{rsi} (line 6). To narrow down the range of \texttt{rdi+0x258}, it is desired that \texttt{rbx} is also controlled by the adversary. We use \texttt{regC} to represent these base registers like \texttt{rbx}.



\bheading{Gadget exploitability.} 
The exploitability of a Type-II gadget is determined by two factors: First, whether there exists a register \texttt{regC} that serves as the base address of the second memory reference. Having such a register makes the attack much easier, because the range of the second memory references can be controlled by the adversary. Second, the number of instructions between the two memory references. Because speculative execution 
only lasts for a very short time, only a few instructions can be executed. The fewer instructions there are in the gadget, the higher its exploitability is. 

\subsection{Symbolically Executing SGX Code}

Although a skillful attacker can manually read the source code or even the disassembled binary code of the enclave program, SGX SDKs, or the runtime libraries to identify usable gadgets for exploitation, such an effort is very tedious and error-prone. It is highly desirable to leverage automated software tools to scan an enclave binary to detect any exploitable gadgets, and eliminate the gadgets before deploying them to the untrusted SGX machines.

To this end, we devise a dynamic symbolic execution technique to enable automated identification of \attackName gadgets. Symbolic execution~\cite{King:1976:symbolic} is a program testing and debugging technique in which symbolic inputs are supplied instead of concrete inputs. Symbolic execution abstractly executes a program and concurrently explores multiple execution paths. The abstract execution of each execution path is associated with a path constraint that represents multiple concrete runs of the same program that satisfy the path conditions. Using symbolic execution techniques, we can explore multiple execution paths in the enclave programs to find gadgets of \attackNames. 

More specifically, we leverage \texttt{angr}~\cite{Shoshitaishvili:2016:state}, a popular binary analysis framework to perform the symbolic execution. During the simulated execution of a program, machine states are maintained internally in \texttt{angr} to represent the status of registers, stacks, and the memory; instructions update the machine states represented with symbolic values while the execution makes forward progress. We leverage this symbolic execution feature of \texttt{angr} to enumerate execution paths and explore each machine state to identify the gadgets.  \looseness=-1





\bheading{Symbolic execution of an enclave function.}
To avoid the path explosion problem during the symbolic execution of a large enclave program (or a large SGX runtime such as Graphene-SGX), we design a tool built atop the \texttt{angr} framework, which allows the user to specify an arbitrary enclave function 
to start the symbolic execution. The exploration of an execution path terminates when the execution returns to this entry function or detects a gadget. 
To symbolically execute an SGX enclave binary, we have extended \texttt{angr} to handle: (1) the \texttt{EEXIT} instruction, by putting the address of the enclave entry point, \texttt{enclave\_entry}, in the \texttt{rip} register of its successor states; 
(2) dealing with instructions that are not already supported by \texttt{angr}, such as \texttt{xsave}, \texttt{xrstore}, \texttt{repz}, and \texttt{rdrand}.   \looseness=-1

\begin{table*}[t]
\scriptsize
\begin{center}
  \begin{tabular}{ | c | p{.18\textwidth} | p{.17\textwidth}  p{.10\textwidth} | p{.34\textwidth} |}
	\hline
     & \textbf{Category} &  \textbf{End Address} & &  \textbf{Controlled Registers} \\
    \hline
    \multicolumn{1}{| c |}{\multirow {11}{*}{Intel SGX SDK}} & indirect jump & $<$do\_ecall$>$:0x118 &  & rdi, r8, r9, r10, r11, r14, r15 \\ \cline{2-5}
    \multicolumn{1}{| c |}{} & indirect call &  ---  &  & --- \\ \cline{2-5}
    \multicolumn{1}{| c |}{} & \multicolumn{1}{ l |}{\multirow {9}{*}{return}} &  $<$get\_enclave\_state$>$:0xc &  & rbx, rdi, rsi, r8, r9, r10, r11, r12, r13, r14, r15 \\ \cline{3-5}
    \multicolumn{1}{| c |}{} & \multicolumn{1}{ c |}{} &  $<$sgx\_is\_enclave\_crashed$>$:0x16 &  & rbx, rdi, rsi, r8, r9, r10, r11, r12, r13, r14, r15 \\ \cline{3-5}
    \multicolumn{1}{| c |}{} & \multicolumn{1}{ c |}{} &  $<$get\_thread\_data$>$:0x9 &  & rbx, rdi, rsi, r8, r9, r10, r11, r12, r13, r14, r15 \\ \cline{3-5}
    \multicolumn{1}{| c |}{} & \multicolumn{1}{ c |}{} &  $<$\_ZL16init\_stack\_guardPv$>$:0x21 &  & rdi, rsi, r8, r9, r10, r11, r12, r13, r14, r15 \\ \cline{3-5}
    \multicolumn{1}{| c |}{} & \multicolumn{1}{ c |}{} &  $<$do\_ecall$>$:0x21 &  & rsi, r8, r9, r10, r11, r12, r13, r14, r15 \\ \cline{3-5}
    \multicolumn{1}{| c |}{} & \multicolumn{1}{ c |}{} &  $<$enter\_enclave$>$:0x62 &  & rbx, rsi, r8, r9, r10, r11, r12, r13, r14, r15 \\ \cline{3-5}
    \multicolumn{1}{| c |}{} &  &  $<$restore\_xregs$>$:0x2b &  & rsi, r8, r9, r10, r11, r12, r14, r15 \\ \cline{3-5}
    \multicolumn{1}{| c |}{} & \multicolumn{1}{ c |}{} &  $<$do\_rdrand$>$:0x11 &  & r8, r9, r10, r11, r12, r14, r15 \\ \cline{3-5}
    \multicolumn{1}{| c |}{} & \multicolumn{1}{ c |}{} &  $<$sgx\_read\_rand$>$:0x46 &  & rbx, r8, r9, r10, r11, r12, r14, r15 \\ \hline 
    \multicolumn{1}{| c |}{\multirow {9}{*}{Graphene-SGX}} & indirect jump & --- &  & --- \\ \cline{2-5}
    \multicolumn{1}{| c |}{} & indirect call & $<$\_DkGenericEventTrigger$>$:0x20  &  & r9, r10, r11, r13, r14, r15 \\ \cline{2-5}
    \multicolumn{1}{| c |}{} & \multicolumn{1}{ l |}{\multirow {7}{*}{return}} & $<$\_DkGetExceptionHandler$>$:0x30  &  & rdi, r8, r9, r10, r11, r12, r13, r14, r15 \\ \cline{3-5}
    \multicolumn{1}{| c |}{} &  &  $<$get\_frame$>$:0x84 &  & r8, r9, r10, r11, r12, r13, r14, r15 \\ \cline{3-5}
    \multicolumn{1}{| c |}{} &   & $<$\_DkHandleExternelEvent$>$:0x55  &  & rdi, r8, r9, r10, r11, r12, r13, r14, r15 \\ \cline{3-5}
    \multicolumn{1}{| c |}{} &  & $<$\_DkSpinLock$>$:0x27  &  & rbx, rdi, r8, r9, r10, r11, r12, r13, r14, r15 \\ \cline{3-5}
    \multicolumn{1}{| c |}{} &  & $<$sgx\_is\_within\_enclave$>$:0x23  &  & rdi, rsi, r8, r12, r13, r14 \\ \cline{3-5}
    \multicolumn{1}{| c |}{} &  & $<$handle\_ecall$>$:0xcd  &  & rdi, rsi, r8 \\ \cline{3-5}
    \multicolumn{1}{| c |}{} &  & $<$handle\_ecall$>$:0xd5  &  & rdx, rdi, rsi, r8 \\ \hline
    \multicolumn{1}{| c |}{\multirow {18}{*}{Rust SGX SDK}} & indirect jump & $<$do\_ecall$>$:0x118 &  & rdi, r9, r10, r11, r12, r13, r14, r15  \\ \cline{2-5}
    \multicolumn{1}{| c |}{} & indirect call & ---  &   &  --- \\ \cline{2-5}
    \multicolumn{1}{| c |}{} & \multicolumn{1}{ l |}{\multirow {15}{*}{return}} & $<$\_ZL14do\_init\_threadPv$>$:0x109  &  & rdi, r9, r10, r11, r12, r13, r14, r15 \\ \cline{3-5}
    \multicolumn{1}{| c |}{} & \multicolumn{1}{ c |}{} & $<$do\_ecall$>$:0x21  &  & rsi, r8, r9, r10, r11, r12, r13, r14, r15 \\ \cline{3-5}
    \multicolumn{1}{| c |}{} & \multicolumn{1}{ c |}{} & $<$do\_ecall$>$:0x63  &  & rsi, r8, r9, r10, r11, r12, r13, r14, r15 \\ \cline{3-5}
    \multicolumn{1}{| c |}{} & \multicolumn{1}{ c |}{} & $<$\_ZL16init\_stack\_guardPv$>$:0x21  &  & rdi, rsi, r8, r9, r10, r11, r12,  r13, r14, r15 \\ \cline{3-5}
    \multicolumn{1}{| c |}{} & \multicolumn{1}{ c |}{} & $<$\_ZL16init\_stack\_guardPv$>$:0x69  &  & rdi, r8, r9, r10, r11, r12, r13, r14, r15 \\ \cline{3-5}
    \multicolumn{1}{| c |}{} & \multicolumn{1}{ c |}{} & $<$enter\_enclave$>$:0x55  &  & rbx, rsi, r8, r9, r10, r11, r12, r13, r14, r15 \\ \cline{3-5}
    \multicolumn{1}{| c |}{} & \multicolumn{1}{ c |}{} & $<$restore\_xregs$>$:0x2b  &  & rsi, r8, r9, r10, r11, r12, r13, r14, r15 \\ \cline{3-5}
    \multicolumn{1}{| c |}{} & \multicolumn{1}{ c |}{} & $<$elf\_tls\_info$>$:0xa0  &  & rbx, rdx, rsi, r9, r10, r11, r14, r15 \\ \cline{3-5}
    \multicolumn{1}{| c |}{} & \multicolumn{1}{ c |}{} & $<$get\_enclave\_state$>$:0xc  &  & rdx, rdi, r8, r9, r10, r11, r12, r14, r15 \\ \cline{3-5}
    \multicolumn{1}{| c |}{} & \multicolumn{1}{ c |}{} & $<$get\_thread\_data$>$:0x9  &  & rbx, rdi, rsi, r8, r9, r10, r11, r12, r13, r14, r15 \\ \cline{3-5}
    \multicolumn{1}{| c |}{} & \multicolumn{1}{ c |}{} & $<$\_\_morestack$>$:0xe  &  & r8, r9, r10, r11 \\ \cline{3-5}
    \multicolumn{1}{| c |}{} & \multicolumn{1}{ c |}{} & $<$asm\_oret$>$:0x64  &  & r8, r9, r10, r11 \\ \cline{3-5}
    \multicolumn{1}{| c |}{} & \multicolumn{1}{ c |}{} & $<$\_\_memcpy$>$:0xa3  &  & rax, rbx, rdi, r9, r10, r11, r14, r15 \\ \cline{3-5}
    \multicolumn{1}{| c |}{} & \multicolumn{1}{ c |}{} & $<$\_\_memset$>$:0x1d  &  & rax, rbx, rdx, rdi, r9, r10, r11, r14, r15 \\ \cline{3-5}
    \multicolumn{1}{| c |}{} & \multicolumn{1}{ c |}{} & $<$\_\_intel\_cpu\_features\_init\_body$>$:0x42b  &  & rbx, rdx, rdi, r9, r10, r11, r14, r15 \\ \hline
  \end{tabular}
\caption{\attackName Type-I Gadgets in Popular SGX Runtime Libraries. 
}
\label{table-gadgets1}
\end{center}
\end{table*}

\subsection{Gadget Identification}

\bheading{Identifying Type-I gadgets.}
 The key requirement of a Type-I gadget is that before the execution of the indirect jump/call or near return instruction, the values of some registers are controlled (directly or indirectly) by the adversary, which can only be achieved via \texttt{EENTER}.  We consider two types of Type-I gadget separately: \texttt{ECall} gadgets and \texttt{ORet} gadgets.


To detect \texttt{ECall} gadgets, the symbolic execution starts from the \texttt{enclave\_entry} function and stops when a Type-I Gadget is found. During the path exploration, \texttt{edi} register is set to a value that leads to an \texttt{ECall}.

To detect \texttt{ORet} gadgets, the symbolic execution starts from a user-specified function inside the enclave. Once an \texttt{OCall} is encountered, the control flow is transfered to \texttt{enclave\_entry} and the \texttt{edi} register is set to a value that leads to an \texttt{ORet}. At this point, all other registers are considered controlled by the adversary and thus are assigned symbolic values. An \texttt{ORet} gadget is found if an indirect jump/call or near return instruction is encountered and some of the registers still have symbolic values. The symbolic execution continues if no gadgets are found until the user-specified function finishes.  \looseness=-1


\bheading{Identifying Type-II gadgets.}
To identify Type-II gadgets, our tool scans the entire enclave binary and looks for memory reference instructions (\ie, \texttt{mov} and its variants, such as \texttt{movd} and \texttt{moveq}) 
that load register \texttt{regB} with data from the memory location pointed to by \texttt{regA}. Both \texttt{regA} and \texttt{regB} are general registers, such as \texttt{rax}, \texttt{rbx}, \texttt{rcx}, \texttt{rdx}, \texttt{r8} - \texttt{r15}. Once one of such instructions is found, the following $N$ instructions (\eg, $N=10$) are examined to see if there exists another memory reference instruction (\eg, \texttt{mov}, \texttt{cmp}, \texttt{add}) that accesses a memory location pointed to by register \texttt{regD}. If so, the instruction sequence is a potential Type-II gadget. It is desired to have a register \texttt{regC} used as the base address for the second memory reference. However, we also consider gadgets that do not involve \texttt{regC}, because they are also exploitable. 

Once we have identified a potential gadget, it is executed symbolically using \texttt{angr}. The symbolic execution starts from the first instruction of a potential Type-II gadget, and \texttt{regB} and \texttt{regC} are both assigned symbolic values. At the end of the symbolic execution of the potential gadget, the tool checks whether \texttt{regD} contains a derivative value of \texttt{regB}, and when \texttt{regC} is used as the base address of the second memory reference, 
whether \texttt{regC} still holds its original symbolic values. The potential gadget is a true gadget if the checks pass. We use either [\texttt{regA}, \texttt{regB}, \texttt{regC}] or [\texttt{regA}, \texttt{regB}] to represent a Type-II gadget.

\subsection{Experimental Results of Gadget Detection}

We run our symbolic execution tool on three well-known SGX runtimes: the official Intel Linux SGX SDK (version $2.1.102.43402$), Graphene-SGX (commit bf90323), and Rust-SGX SDK (version $0.9.1$). In all cases, a minimal enclave with a single empty \texttt{ECall} was developed. When the enclave binary becomes more complex (\eg, using some library functions such as \texttt{printf}), the size of the resulting enclave binary will grow to include more components of the SDK libraries. Therefore, gadgets detected in a minimal enclave binary will appear in any enclave code developed using these SDKs; Additional functionality will increase the number of available gadgets. For example, a simple \texttt{OCall} implementation of \texttt{printf} introduces three more Type-II gadgets. In addition, the code written by the enclave author might also introduce extra exploitable gadgets.

To detect \texttt{ECall} Type-I Gadgets, the symbolic execution starts from the \texttt{enclave\_entry} function in all three runtime libraries. To detect \texttt{ORet} Type-I gadgets, in Intel SGX SDK and Rust-SGX SDK, we started our analysis from the \texttt{sgx\_ocall} function, which is the interface defined to serve all \texttt{OCalls}. In contrast, Graphene-SGX has more diverse \texttt{OCalls} sites. In total, there are 37 such sites as defined in \texttt{enclave\_ocalls.c}. Unlike in other cases where the symbolic analysis completes instantly due to small function sizes, analyzing these 37 \texttt{OCalls} sites consumes more time: the median running time of analyzing one \texttt{OCalls} sites was 39 seconds; the minimum analysis time was 8 seconds; and the maximum was 340 seconds.  \looseness=-1

The results for Type-I gadgets are summarized in \tabref{table-gadgets1} and those for Type-II gadgets are listed in \tabref{table-gadgets2}. More specifically, in \tabref{table-gadgets1}, column 2 shows the type of the gadget, whether it being \textit{indirect jump}, \textit{indirect call}, or \textit{return}; column 3 shows the  address of the branch instruction (basically the gadget's end address. Note that the Type-I gadget always starts at the \texttt{enclave\_entry}.) represented using the function name the instruction is located and its offset; column 4 shows the registers that are under the control of the adversary when the branch instructions are executed. For example, the first entry in \tabref{table-gadgets1} shows an indirect jump gadget, which is located in \texttt{do\_ecall} (with an offset of 0x118). By the time of the indirect jump, the registers that are still under the control of adversary are \texttt{rdi}, \texttt{r8}, \texttt{r9}, \texttt{r10}, \texttt{r11}, \texttt{r14} and \texttt{r15}.

\tabref{table-gadgets2} (in Appendix) lists Type-II gadgets of the form [\texttt{regA}, \texttt{regB}, \texttt{regC}], which means at the time of memory reference, two registers, \texttt{regB} and \texttt{regC}, are controlled by the adversary. Such gadgets are easier to exploit. Column 2 shows the beginning address of the gadgets, represented using the function name and offset within the function; column 3 lists the entire gadgets. For most of these gadgets, the number of instructions in the gadget is less than 5. The shorter the gadgets are, the easier they can be exploited. The Type-II gadgets of the form [\texttt{regA}, \texttt{regB}] were not listed in the table, because there are too many. In total, we have identified $6$, $86$, and $180$ such gadgets in these three runtimes, respectively.



\section{Stealing Enclave Secrets with \attackNames}
\label{sec:exploit}

\begin{figure}[t]
    \includegraphics[width=0.5\textwidth]{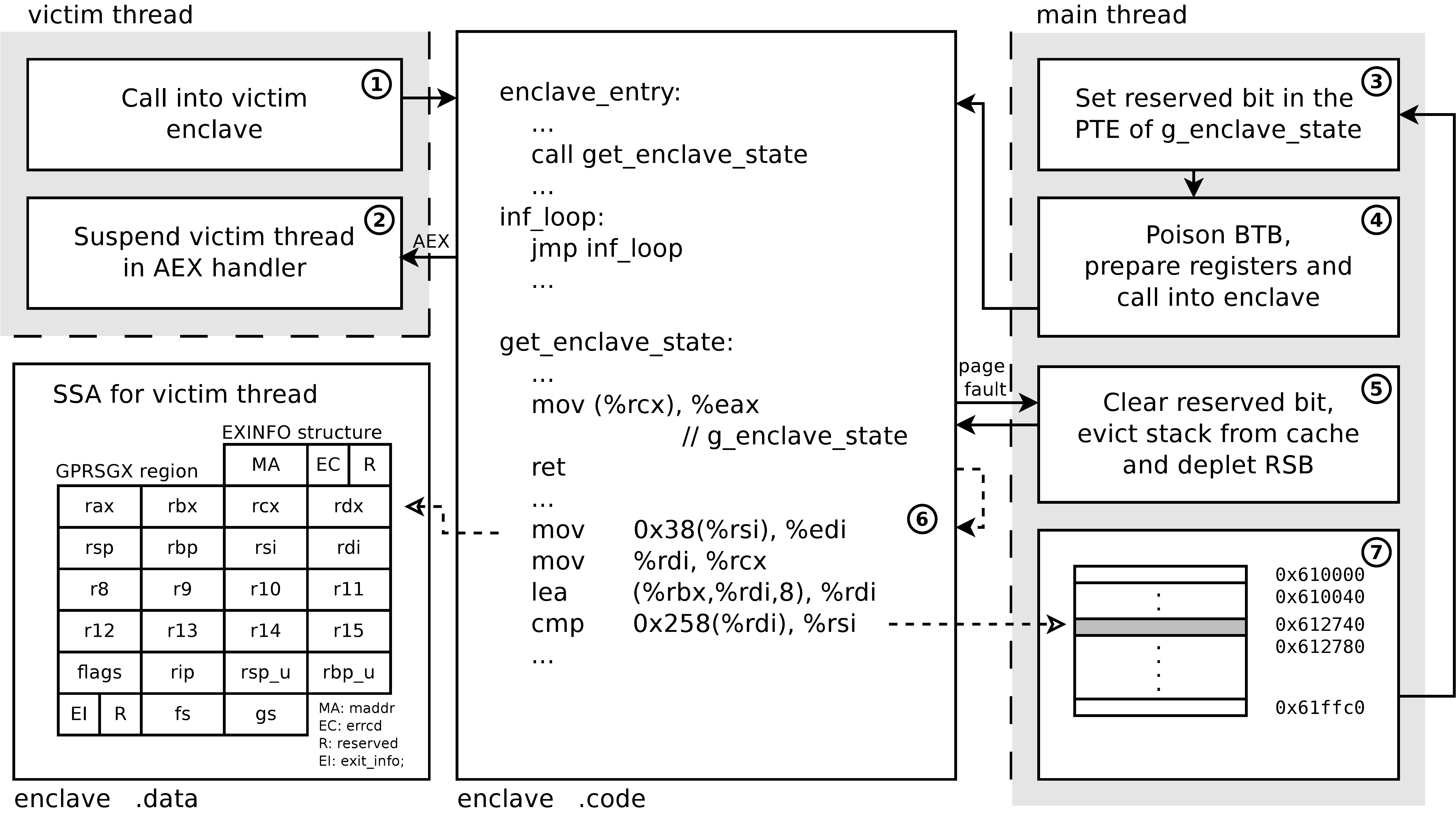}
  \vspace{-0.1in}
   \caption{Exploiting Intel SGX SDK. The blocks with dark shadows represent instructions or data located in untrusted memory. Blocks without shadows are instructions inside the target enclave or the \texttt{.data} segment of the enclave memory.}
   \label{fig:exploit}
  \vspace{-0.1in}
\end{figure}


In this section, we demonstrate end-to-end \attackNames against an arbitrary enclave program written with Intel SGX SDK~\cite{sgxsdk}, because this is Intel's official SDK. Rust-SGX was developed based on the official SDK and thus can be exploited in the same way. For demonstration purposes, the enclave program we developed has only one \texttt{ECall} function that runs in a busy loop. We verified that our own code does not contain any Type-I or Type-II gadgets in itself. The exploited gadgets, however, are located in the runtime libraries of SDK version $2.1.102.43402$ (compiled with \texttt{gcc} version $5.4.0 20160609$), which are listed in \autoref{lst:gadgetI} and \autoref{lst:gadgetII}. Experiments were conducted on a Lenovo Thinkpad X1 Carbon ($4$th Gen) laptop with an Intel Core i$5$-$6200$U processor and $8$GB memory. 

%


\subsection{Reading Register Values}
\label{sec:exploit:ssa}
We first demonstrate an attack that enable the adversary to read arbitrary register values inside the enclave. This attack is possible because during AEX, the values of registers are stored in the SSA before exiting the enclave. As the SSA is also a memory region inside the enclave, the adversary could leverage the \attackNames to read the register values in the SSA during an AEX. This attack is especially powerful as it allows the adversary to frequently interrupt the enclave execution with AEX~\cite{VanBulck:2017:SPA:3152701.3152706} and take snapshots of its SSAs to single-step trace its register values during its execution.

In particular, the attack is shown in \autoref{fig:exploit}. In \st{192},
the targeted enclave code is loaded into the enclave that is created by the program controlled by the adversary. After \texttt{EINIT}, the malicious program starts a new thread (denoted as the victim thread) to issue \texttt{EENTER} to execute the enclave code. Our enclave code only runs in a busy loop. But in reality, the enclave program might complete a remote attestation and establish trusted communication with its remote owner. In \st{193}, the adversary triggers frequent interrupts to cause AEX from the targeted enclave. During an AEX, the processor stores the register values into the SSA , exits the enclave and invokes the system software's interrupt handler. Before the control is returned to the enclave program via \texttt{ERESUME}, the adversary pauses the victim thread's execution at the AEP, a piece of instructions in the untrusted runtime library that takes control after \texttt{IRet}. 

In \st{194}, the main thread of the adversary-controlled program sets (through a kernel module) the reserved bit in the PTE of an enclave memory page that holds  \texttt{g\_enclave\_state}, a global variable used by Intel SGX SDK to track the state of the enclave, \eg,  initialized or crashed states.  As shown in \lstref{lst:gadgetI}, this global variable is accessed right before the \texttt{ret} instruction of the Type-I gadget (\ie, the memory referenced by \texttt{rcx} in the instruction ``\texttt{mov (\%rcx),\%eax}''. In \st{195}, the main thread poison the BTB, prepare registers (\ie, \texttt{rsi} and \texttt{rdi}\footnote{Note that \texttt{rbx} will be set to \texttt{rdi} by the time the return instruction is executed (line 34 in \autoref{lst:gadgetI}), in such a way we can control \texttt{rsi} and \texttt{rbx} when speculatively executing Type-II gadget.}), and executes \texttt{EENTER} to trigger the attack. 
To poison the BTB, the adversary creates an auxiliary enclave program in another process containing an indirect jump with the source address equals the address of the return instruction in the Type-I gadget, and the target address the same as the start address of the Type-II gadget in the victim enclave. The process that runs in the auxiliary enclave is pinned onto the same logical core as the main thread. To trigger the BTB poisoning code, the main thread calls \texttt{sched\_yield()} to relinquish the logical core to the auxiliary enclave program.

In \st{196}, after the main thread issues \texttt{EENTER} to get into the enclave mode, the Type-I gadget will be executed immediately. Because a reserved bit in the PTE is set, a page fault is triggered when the enclave code accesses the global variable \texttt{g\_enclave\_state}. In the page fault handler, the adversary clears the reserved bit in the PTE, evicts the stack frame that holds the return address of the \texttt{ret} instruction from cache by accessing $2,000$ memory blocks whose virtual addresses have the same lower $12$-bits as the stack address. The RSB is depleted right before \texttt{ERESUME} from the fault handling, so that it will remain empty until the return instruction of Type-I gadget is executed.
In \st{197}, due to the extended delay of reading the return address from memory, the processor speculatively executes the Type-II gadget (as a result of the BTB poisoning and RSB depletion).  After the processor detects the mis-prediction and flushes the speculatively executed instructions from the pipeline, the enclave code continues to execute. However, because \texttt{rdi} is set as a memory address in our attack, 
it is an invalid value for the SDK as \texttt{rdi} is used as the index of the \texttt{ecall\_table}. The enclave execution will return with an error quickly after the speculative execution. This artifact allows the adversary to repeatedly probe into the enclaves. In \st{198}, the adversary uses \flushreload techniques to infer the memory location accessed inside the Type-II gadget. One byte of SSA can thus be leaked. The main thread then repeats \st{194} to \st{198} to extract the remaining bytes of the SSA. \looseness=-1

In our Type-I gadget, the \texttt{get\_enclave\_state} function is very short as it contains only $4$ instructions. Since calling into this function will load the stack into the L1 cache, it is very difficult to flush the return address out of the cache to win the race condition. In fact, our initial attempts to flush the return address all failed. Triggering page faults to flush the return address resolves the issue. However, directly introducing page faults in every stack access could greatly increase the amount of time to carry out the attack. Therefore, instead of triggering page faults on the stack memory, the page fault is enforced on the global variable \texttt{g\_enclave\_state} which is located on another page. In this way, we can flush the return address with only one page fault in each run. 

In our Type-II gadget, the first memory access reads $4$ bytes ($32$ bits).  It is unrealistic to monitor $2^{32}$ possible values in the \flushreload. However, if we know the value of lower 24 bits, we can adjust the base of the second memory access (\ie, \texttt{rbx}) to map the $256$ possible values of the highest 8 bits to the cache lines monitored by the \flushreload code. 
Once all 32 bits of the targeted memory are learned, the adversary shifts the target address by one byte to learn the value of a new byte.  
We found in practice that it is not hard to find the initial consecutively known bytes. For example, the unused bytes in an enclave data page will be initialized as $0$x$00$, as they are used to calculate the measurement hash. Particularly, we found that there are $4$ reserved bytes (in the EXINFO structure) in the SSA right before the GPRSGX region (which stores registers).
Therefore, we can start from the reserved bytes (all $0$s), and extract the GPRSGX region from the first byte to the last.
As shown in \figref{fig:exploit}, all register values, including \texttt{rax},  \texttt{rbx}, \texttt{rcx}, \texttt{rdx}, \texttt{r8} to \texttt{r15}, \texttt{rip}, \etc, can be read from the SSA very accurately. To read all registers in the GPRSGX region ($184$ bytes in total), our current implementation takes $414$ to $3677$ seconds to finish. On average, each byte can be read in $6.6$ seconds. We believe our code can be further improved.


\subsection{Stealing Intel Secrets}
\label{subsec:readmem}

Reading other enclave memory follows exactly the same steps. The primary constraint is that the attack is much more convenient if three consecutive bytes are known. To read the \texttt{.data} segments, due to data alignment, some bytes are reserved and initialized as $0$s, which can be used to bootstrap the attack. In addition, some global variables have limited data ranges, rendering most bytes known. To read the stack frames, the adversary could begin with a relatively small address which is likely unused and thus is known to be initialized with $0$x$cc$. In this way, the adversary can start reading the stack frames from these known bytes. Next, we demonstrate how to use these memory reading primitives to steal Intel secrets, such as seal keys and attestation keys. 

\bheading{Extracting seal keys and decrypting sealed storage blob.}
The adversary could use \attackNames to read the seal keys from the enclave memory when it is being used during sealing or unsealing operations. Particularly, in our demonstration, we targeted Intel SDK API \texttt{sgx\_unseal\_data()} used for unsealing a sealed blob. The \texttt{sgx\_unseal\_data()} API works as follows: firstly, it calls \texttt{sgx\_get\_key()} function to generate the seal key from a pseudo-random function inside the processor and then store it temporarily on the stack in the enclave memory. Secondly, with the seal key, it calls \texttt{sgx\_rijndael128GCM\_decrypt()} function to decrypt the sealed blob. Finally, it clears the seal key (by setting the memory range storing the seal key on the stack to $0$s) and returns. Hence, to read the seal key, the adversary suspends the execution of the victim enclave when function \texttt{sgx\_rijndael128GCM\_decrypt()} is being called, by setting the reserved bit of the PTE of the enclave code page containing \texttt{sgx\_rijndael128GCM\_decrypt()}. The adversary then launches the \attackNames to read the stack and extract the seal key.

To decrypt the sealed blob, the adversary could exported the seal key and then implement the AES-128-GCM decryption algorithm by himself to decrypt the sealed blob with the seal key. This may happen outside the enclave or on a different machine, because the SGX hardware is no longer involved in the process.

\bheading{Extracting attestation key}.
After running the provisioning protocol with Intel's provisioning service, the attestation key (\ie, EPID private key) is created and then sealed in the EPID blob by the provisioning enclave and stored on a non-volatile memory. Though the location of the non-volatile memory is not documented, during remote attestation, SGX still relies on the untrusted OS to pass the sealed EPID blob into the quoting enclave. This offers the adversary a chance to obtain the sealed EPID blob.

To decrypt the EPID blob and extract the attestation key, the adversary could target the \texttt{verify\_blob()} \texttt{ECall} function of the quoting enclave which is used to verify the sealed EPID blob, suspend its execution when \texttt{sgx\_rijndael128GCM\_decrypt()} is being called, and read the stack to obtain the quoting enclave's seal key. With this seal key, the attestation key can be decrypted in similar ways as the aforementioned attack.

The primary difference in this attack is that it requires the adversary to perform \attackNames on Intel signed quoting enclaves, rather than ISV's enclaves. For a real attacker, these two attacks are similar. But it made a big difference in our experiments as Intel's enclaves are developed and signed by Intel, which cannot be altered. Nevertheless, we could perform the attack using the same method described in the paper. One thing worth mentioning is that the TCS numbers of the provisioning enclave and quoting enclave are set to $1$, which means the adversary has to use the same TCS to enter the enclaves. Since the number of SSAs per TCS is $2$, which is designed to allow the victim to run some exception handler within the enclave when the exception could not be resolved outside the enclave during AEXs. However, this also enables the adversary to \texttt{EENTER} into the enclave during an AEX, to launch the \attackName to steal the secrets being used by the victim.

After the attestation key is obtained, the adversary could use this EPID private key to generate an anonymous group signature, which means the adversary can now impersonate any machine in the attestation group. Moreover, the adversary could also use the attestation key completely outside the enclave and trick the ISVs to believe their code runs inside an enclave.

\section{Countermeasures}
\label{sec:counter}



\bheading{Hardware patches.}
To mitigate branch target injection attacks, Intel has released microcode updates to support the following three features~\cite{IntelSpectre}. 

\begin{packeditemize}

\item \textit{Indirect Branch Restricted Speculation (IBRS):} 
IBRS restricts the speculation of indirect branches~\cite{IntelSpectreMitigate}. Software running in a more privileged mode can set an architectural model-specific register (MSR), \texttt{IA32\_SPEC\_CTRL.IBRS}, to 1 by using the \texttt{WRMSR} instruction, so that indirect branches will not be controlled by software that was executed in a less privileged mode or by a program running on the other logical core of the physical core. By default, on machines that support IBRS, branch prediction inside the SGX enclave cannot be controlled by software running in the non-enclave mode. 



\item \textit{Single Thread Indirect Branch Predictors (STIBP):}
STIBP prevents branch target injection from software running on the neighboring logical core, which can be enabled by setting \texttt{IA32\_SPEC\_\linebreak[0]CTRL.STIBP} to 1 by using the \texttt{WRMSR} instruction.

\item \textit{Indirect Branch Predictor Barrier (IBPB):}
IBPB is an indirect branch control command that establishes a barrier to prevent the branch targets after the barrier from being controlled by software executed before the barrier. The barrier can be established by setting the \texttt{IA32\_PRED\_CMD.IBPB} MSR using the \texttt{WRMSR} instruction.

\end{packeditemize}

Particularly, IBPS provides a default mechanism that prevents branch target injection. To validate the claim, we developed the following tests: First, to check if the BTB is cleansed during \texttt{EENTER} or \texttt{EEXIT}, 
we developed a dummy enclave code that trains the BTB to predict address $A$ for an indirect jump. After training the BTB, the enclave code uses  \texttt{EEXIT} and a subsequent \texttt{EENTER} to switch the execute mode once and then executes the same indirect jump but with address $B$ as the target. Without the IBRS patch, the later indirect jump will speculatively execute instructions in address $A$. However, with the hardware patch, instructions in address $A$ will not be executed. 

Second, to test if the BTB is cleansed during \texttt{ERESUME}, we developed another dummy enclave code that will always encounter an AEX (executing a memory access to a specific address that will trigger a page fault) right before an indirect call. In the AEP, another BTB poisoning enclave code will be executed before \texttt{ERESUME}. Without the patch, the indirect call speculatively executed the secret-leaking gadget. The attack failed after patching.

Third, to test the effectiveness of the hardware patch under Hyper-Threading, we tried poisoning the BTB using a program running on the logical core sharing the same physical core. The experiment setup was similar to our end-to-end case study in \secref{sec:exploit}, but instead of pinning the BTB poisoning enclave code onto the same logical core, we pinned it onto the sibling logical core. We observed some secret bytes leaked before the patch, but no leakage after applying the patch.

Therefore, from these tests, we can conclude that SGX machines with microcode patch will cleanse the BTB during \texttt{EENTER} and during \texttt{ERESUME}, and also prevent branch injection via Hyper-Threading, thus they are immune to \attackNames.

\bheading{Retpoline.}
Retpoline is a pure software-based solution to Spectre attacks~\cite{retpoline}, which has been developed for major compilers, such as GCC~\cite{retpoline:gcc} and LLVM~\cite{retpoline:llvm}. The name ``retpoline'' comes from ``return'' and ``trampoline''. Because modern processors have implemented separate predictors for function returns, such as Intel's return stack buffer~\cite{US5964868, US6170054, US6253315, US6374350, US6560696} and AMD's return-address stack~\cite{Keltcher:2003:AOP}, it is believed that these return predictors are not vulnerable to Spectre attacks. Therefore, the key idea of retpoline is to replace indirect jump or indirect calls with returns to prevent branch target injection.

However, in recent Intel Skylake/Kabylake processors, on which SGX is supported, when the RSB is depleted, the BPU will fall back to generic BTBs to predict a function return. This allows poisoning of return instructions. Therefore, Retpoline is useless by itself in preventing \attackNames. 

\bheading{Defenses by Intel's attestation service.} After applying the microcode patch, the processor is immune to \attackNames. But unpatched processors remain vulnerable. The key to the security of the SGX ecosystem is whether attestation measurements and signatures from processors without the IBRS patch can be detected during remote attestation. As discussed in \secref{sec:background}, the \texttt{CPUSVN} is used to derive attestation keys (indirectly) and seal keys, and also provided to the attestation service; microcode update will also upgrade \texttt{CPUSVN}. As a consequence, any attestation key and seal key generated before the microcode update will not be trustworthy afterwards. Moreover, Intel's attestation service, which arbitrates every attestation request from the ISV, responses to the attestation signatures generated from unpatched CPUs with an error message indicating outdated \texttt{CPUSVN}.

\bheading{Summary.}
The combination of the IBPS patch and defenses by Intel's attestation service has been an effective defense against \attackNames. However, there are several caveats:
First, any secret that is allowed to be provisioned to an unpatched processor can be leaked. This includes secrets in ISV enclaves that are provisioned before remote attestation, or after remote attestation if the ISV chooses to ignore the error message returned by the attestation service. Moreover, because the ISV enclave's seal key can be compromised by \attackNames, any secret sealed by an enclave run on unpatched processor can be decrypted by the adversary. Furthermore, any legacy sealed secrets become untrustworthy, as they could be forged by the adversary using the stolen seal key.

Second, as shown in \secref{subsec:readmem}, the EPID private key used in the remote attestation can be extracted by the attacker. Given the anonymous attestation protocol~\cite{intelepid} used by Intel, the attacker can provide a valid signature for any SGX processors in the group. With the attestation key, it is also possible for the attacker to run the enclave code entirely outside the enclave and forge a valid signature to fool the ISV. Therefore, an error message during attestation with \texttt{GROUP\_OUT\_OF\_DATE} means the enclave is completely untrusted, rather than replying on ISV to decide (see \tabref{table-attestation}). We recommend Intel to make the message very clear to the ISVs.

\begin{table}[]
\scriptsize
\centering

\begin{tabular}{|p{.147\textwidth}|p{.215\textwidth}|p{.057\textwidth}|}
\hline
Result                   & Description                                                                                & Trustworthy \\ \hline
OK                       & EPID signature was verified correctly and the TCB level of the SGX platform is up-to-date. & Yes                                  \\ \hline
SIGNATURE\_INVALID       & EPID signature was invalid.                                                                & No                                   \\ \hline
GROUP\_REVOKED           & EPID group has been revoked.                                                               & No                                   \\ \hline
SIGNATURE\_REVOKED       & EPID private key used has been revoked by signature.                                       & No                                   \\ \hline
KEY\_REVOKED             & EPID private key used has been directly revoked (not by signature).                        & No                                   \\ \hline
SIGRL\_VERSION\_MISMATCH & SigRL version does not match the most recent version of the SigRL.                         & No                                   \\ \hline
GROUP\_OUT\_OF\_DATE     & EPID signature was verified correctly, but the TCB level of SGX platform is outdated.      & Up to ISV           \\ \hline
\end{tabular}
\caption{Attestation Results~\cite{inteliasapi}}
\label{table-attestation}
\end{table}

Due to the severity of \attackNames, we urge the enclave authors to specify the minimum \texttt{CPUSVN} during their development. It is important never accept attestation from processors with outdated \texttt{CPUSVN}. Moreover, we also suggest developers of runtime libraries (such as SGX SDKs) to scrutinize their code to remove exploitable gadgets in prevention of other potential ways of poisoning the BTB in the future. The symbolic execution tool presented in this paper can be used to look for these gadgets. Type-II gadgets can be removed by adding \texttt{lfense} in between of the two memory references. But the performance loss needs to be evaluated as [\texttt{regA}, \texttt{regB}] Type-II gadgets are very common in the runtimes. Type-I gadgets are harder to be eliminated, as it requires almost all registers to be sanitized after \texttt{EENTER} and before the control flows reach any indirect branch instructions or near returns. 


\section{Related Work}
\label{sec:related}


\bheading{Meltdown and Spectre attacks.}
Our work is closely related to the recently demonstrated Spectre attacks~\cite{spectre, googleprojectzero}. There are two variants of Spectre attacks: bounds check bypass and branch target injection. The first variant targets the conditional branch prediction and the second targets the indirect jump target prediction. A variety of attack scenarios have been demonstrated, including cross-process memory read~\cite{spectre}, kernel memory read from user process, and host memory read from KVM guests~\cite{googleprojectzero}. However, their security implications on SGX enclaves have not been studied. In contrast, in this paper we have systematically investigated the enclave security on vulnerable SGX machines, devised new techniques to enable attacks against any enclave programs developed with Intel SGX SDK, and examined the effectiveness of various countermeasures. 

Meltdown attacks~\cite{meltdown} are another micro-architectural side-chan-nel attacks that exploit implicit caching to extract secret memory content that is not directly readable by the attack code. Different from Spectre attacks, Meltdown attacks leverage the feature of out-of-order execution to execute instructions that should  have not been executed. An example given by Lipp \etal~\cite{meltdown} showed that an unprivileged user program could access an arbitrary kernel memory element and then visit a specific offset in an attacker-controlled data array, in accordance with the value of the kernel memory element, to load data into the cache. Because of the out-of-order execution, instructions after the illegal kernel memory access can be executed and then discarded when the kernel memory access instruction triggers an exception. However, due to implicit caching, the access to the attacker-controlled data array will leave traces in the cache, which will be captured by subsequent \flushreload measurements. Similar attacks can be performed to attack Xen hypervisor when the guest VM runs in paravirtualization mode~\cite{meltdown}. 
However, we are not aware of any demonstrated Meltdown attacks against SGX enclaves. 

\bheading{Micro-architectural side channels in SGX.}
The \attackNames are variants of micro-architectural side-channel attacks. Previously, various micro-architectural side-channel attacks have been demonstrated on SGX, which CPU cache attacks~\cite{Schwarz:2017:MGE, Brasser:2017:SGE, hahnel:2017:HRS, Gotzfried:2017:CAI}, BTB attacks~\cite{Lee:2017:IFC}, page-table attacks~\cite{Xu:2015:CAD, Shinde:2015:PYF, Van:2017:TYS}, cache-DRAM attacks~\cite{Wang:2017:LCD}, \etc.  \attackNames are different because they target memory content inside enclaves, while previous attacks aim to learn secret-dependent memory access patterns. However, \attackNames leverage techniques used in these side-channel attacks to learn ``side effects'' of speculatively executed enclave code.


\bheading{Side-channel defenses.} 
Existing countermeasures to side-channel attacks can be categorized into three classes: hardware solutions, system solutions, and application solutions. Hardware solutions~\cite{Wang:2006:CSC, Wang:2007:NCD, Domnitser:2012:NCL, Martin:2012:TRT, Liu:2014:RFC, Costan:2016:sanctum} require modification of the processors, which are typically effective, but are limited in that the time window required to have major processor vendors to incorporate them in commercial hardware is very long. 
System solutions only modify system software~\cite{Kim:2012:SSP, Varadarajan:2014:SDC, Liu:2016:catalyst, Zhou:2016:SAD}, but as they require trusted system software, they cannot be directly applied to SGX enclaves. 

Application solutions are potentially applicable to SGX. Previous work generally falls into three categories: First, using compiler-assisted approaches to eliminate secret-dependent control flows and data flows~\cite{Molnar:2005:PCS, Coppens:2009:PMT, Shinde:2015:PYF}, or to diversify or randomize memory access patterns at runtime to conceal the true execution traces~\cite{Crane:2015:diversity, Rane:2015:raccoon}. However, as the vulnerabilities in the enclave programs that enable \attackNames are not caused by secret-dependent control or data flows, these approaches are not applicable.
Second, using static analysis or symbolic execution to detect cache side-channel vulnerabilities in commodity software~\cite{Doychev:2013:CTS, Wang:2017:cached}. However, these approaches model secret-dependent memory accesses in a program; they are not applicable in the detection of the gadgets used in our attacks. 
Third, detecting page-fault attacks or interrupt-based attacks against SGX enclave using Intel's hardware transactional memory~\cite{Shih:2017:tsgx,Chen:2017:dejavu, Fu:2017:sgx-lapd}. These approaches can be used to detect frequent AEX, but still allowing secret leaks in  \attackNames. 
\section{Conclusion}
\label{sec:conclude}

We have presented \attackNames that are able to extract the Intel secrets such as the seal keys and attestation keys from the SGX enclaves. To demonstrate their practicality, we systematically explored the possible vectors of branch target injection, approaches to win the race condition during enclave's speculative execution, and techniques to automatically search for code patterns required for launching the attacks. We also demonstrated a number of practical attacks against an arbitrary enclave program written with Intel SGX SDK, which not only extracts the secrets in the enclave memory, but also the registers used only in the enclave mode. 



\section*{Acknowledgments}
\label{sec:ack}
The work was supported in part by the NSF grants 1566444, 1718084, 1750809, 1834213, and 1834215.

\bibliographystyle{ACM-Reference-Format}
\bibliography{paper}

\section{Appendix}
\label{sec:appendix}


Due to space constraints, we list all [\texttt{regA}, \texttt{regB}, \texttt{regC}] Type-II gadgets of the three SGX runtimes, \eg, Intel SGX SDK, Graphene-SGX, and Rust-SGX SDK, in \tabref{table-gadgets2}. The numbers of [\texttt{regA}, \texttt{regB}] Type-II gadgets are too large to be included in the paper.

\begin{table*}[t]
\scriptsize

\begin{center}
  \begin{tabular}{ | c |  p{.24\textwidth} | p{.58\textwidth} |}
	\hline
	&  \textbf{Start Address} &  \textbf{Gadget Instructions} \\ \hline
    \multicolumn{1}{| c |}{\multirow {6}{*}{Intel SGX SDK}}  & $<$dispose\_chunk$>$:0x8a  & mov 0x38(\%rsi),\%r9d; mov \%r9,\%rcx; lea (\%rdi,\%r9,8),\%r9; cmp 0x258(\%r9),\%rsi   \\ \cline{2-3}
    \multicolumn{1}{| c |}{}  &  $<$dispose\_chunk$>$:0x299 & mov 0x38(\%r8),\%r9d; mov \%r9,\%rcx; lea (\%rdi,\%r9,8),\%r9; cmp 0x258(\%r9),\%r8  \\ \cline{2-3}
    \multicolumn{1}{| c |}{}  & $<$dlmalloc$>$:0x180b  & mov 0x38(\%rdx),\%r12d; mov \%r12,\%rcx; add \$0x4a,\%r12; cmp 0x8(\%rsi,\%r12,8),\%rdx  \\ \cline{2-3}
    \multicolumn{1}{| c |}{}   & $<$dlfree$>$:0x399  & mov 0x38(\%r8),\%edi; mov \%rdi,\%rcx; lea (\%rbx,\%rdi,8),\%rdi; cmp 0x258(\%rdi),\%r8  \\ \cline{2-3} 
    \multicolumn{1}{| c |}{}   & $<$dlfree$>$:0x46f  & mov 0x38(\%rsi),\%edi; mov \%rdi,\%rcx; lea (\%rbx,\%rdi,8),\%rdi; cmp 0x258(\%rdi),\%rsi  \\ \cline{2-3} 
    \multicolumn{1}{| c |}{}  & $<$dlrealloc$>$:0x341  & mov 0x38(\%rsi),\%r10d;mov \%r10,\%rcx;lea (\%rbx,\%r10,8),\%r10;cmp \%rsi,0x258(\%r10)  \\ \hline 
    \multicolumn{1}{| c |}{\multirow {20}{*}{Graphene-SGX}}  & $<$do\_lookup\_map$>$:0x97  & mov 0x2f0(\%r8),\%rax; mov  (\%rax,\%rdx,4),\%eax \\ \cline{2-3}
    \multicolumn{1}{| c |}{}   & $<$do\_lookup\_map$>$:0x177  & mov 0x2d0(\%r8),\%rax; mov (\%rax,\%rdx,4),\%r15d \\ \cline{2-3}
    \multicolumn{1}{| c |}{}   & $<$do\_lookup\_map$>$:0x200  & mov 0x2d8(\%r8),\%rax; mov (\%rax,\%r15,4),\%r15d \\ \cline{2-3}
    \multicolumn{1}{| c |}{}   & $<$mbedtls\_mpi\_safe\_cond\_assign$>$:0x98  & mov 0x10(\%r12),\%rcx; movslq \%r9d,\%rdi; mov \%rdi,\%rsi; imul (\%rcx,\%rdx,8),\%rsi \\ \cline{2-3}
    \multicolumn{1}{| c |}{}   & $<$mbedtls\_mpi\_get\_bit$>$:0x13  & mov 0x10(\%rdi),\%rax;mov \%rsi,\%rdx;mov \%esi,\%ecx;shr \$0x6,\%rdx;mov (\%rax,\%rdx,8),\%rax \\ \cline{2-3}
    \multicolumn{1}{| c |}{}   & $<$mbedtls\_mpi\_set\_bit$>$:0x32  & mov 0x10(\%r12),\%rax; mov \%r13,\%rcx; and \$0x3f,\%ecx; shl \%cl,\%rbx; lea (\%rax,\%r14,8),\%rdx; mov \$0xfffffffffffffffe,\%rax; rol \%cl,\%rax; and (\%rdx),\%rax \\ \cline{2-3}
    \multicolumn{1}{| c |}{}   & $<$mbedtls\_mpi\_shift\_l$>$:0x4a  & mov 0x10(\%r13),\%rdx; sub \%rbx,\%rax; lea (\%rdx,\%rax,8),\%rax; mov -0x8(\%rax),\%rcx \\ \cline{2-3}
    \multicolumn{1}{| c |}{}   & $<$mbedtls\_mpi\_shift\_l$>$:0x8a  & mov 0x10(\%r13),\%rsi; mov \$0x40,\%edi; mov \%r12d,\%r8d; sub \%r12d,\%edi; xor \%eax,\%eax; mov (\%rsi,\%rbx,8),\%rdx \\ \cline{2-3}
    \multicolumn{1}{| c |}{}   & $<$mbedtls\_mpi\_cmp\_abs$>$:0x7c  & mov 0x10(\%rdi),\%rax; mov -0x8(\%rax,\%rdx,8),\%rdi \\ \cline{2-3}
    \multicolumn{1}{| c |}{}   & $<$mpi\_montmul.isra.3$>$:0xa0  & mov 0x10(\%r15),\%rdx; mov (\%r14),\%rsi; mov -0x58(\%rbp),\%rdi; mov (\%rdx,\%r13,8),\%r8 \\ \cline{2-3}
    \multicolumn{1}{| c |}{}   & $<$mbedtls\_mpi\_cmp\_mpi$>$:0x91  & mov 0x10(\%rdi),\%rcx; mov -0x8(\%rcx,\%rdx,8),\%rsi \\ \cline{2-3}
    \multicolumn{1}{| c |}{}   & $<$mbedtls\_mpi\_mul\_mpi$>$:0x100  & mov 0x10(\%r13),\%rax; mov \%r11,\%rdx; add 0x10(\%rbx),\%rdx; mov 0x10(\%r12),\%rsi; mov \%r14,\%rdi; mov (\%rax,\%r11,1),\%rcx \\ \cline{2-3}
    \multicolumn{1}{| c |}{}   & $<$mbedtls\_mpi\_mod\_int$>$:0x37  & mov 0x10(\%rsi),\%r11; xor \%ecx,\%ecx; mov -0x8(\%r11,\%r10,8),\%r9 \\ \cline{2-3}
    \multicolumn{1}{| c |}{}   & $<$mbedtls\_mpi\_write\_string$>$:0x129  & mov 0x10(\%r14),\%rax; lea 0x0(,\%rdx,8),\%ecx; mov (\%rax,\%r8,1),\%rax \\ \cline{2-3}
    \multicolumn{1}{| c |}{}   & $<$mbedtls\_aes\_setkey\_enc$>$:0x108  & mov 0xc(\%rbx),\%edi; add \$0x4,\%r8; add \$0x10,\%rbx; mov \%rdi,\%rdx; movzbl \%dh,\%edx; movzbl (\%rsi,\%rdx,1),\%ecx \\ \cline{2-3}
    \multicolumn{1}{| c |}{}   & $<$mbedtls\_aes\_setkey\_enc$>$:0x1e8 & mov 0x1c(\%rbx),\%r8d; add \$0x20,\%rbx; add \$0x4,\%rdi; mov \%r8,\%rdx; movzbl \%dh,\%edx; movzbl (\%rsi,\%rdx,1),\%r9d \\ \cline{2-3}
    \multicolumn{1}{| c |}{}   & $<$mbedtls\_aes\_setkey\_enc$>$:0x238  & mov -0x14(\%rbx),\%edx; mov \%ecx,(\%rbx); xor -0x1c(\%rbx),\%ecx; mov \%ecx,0x4(\%rbx); xor -0x18(\%rbx),\%ecx; xor \%ecx,\%edx; mov \%ecx,0x8(\%rbx); movzbl \%dl,\%ecx; mov \%edx,0xc(\%rbx); movzbl (\%rsi,\%rcx,1),\%r9d \\ \cline{2-3}
    \multicolumn{1}{| c |}{}   & $<$mbedtls\_aes\_setkey\_enc$>$:0x2c8  & mov 0x14(\%rbx),\%edi; add \$0x18,\%rbx; add \$0x4,\%r8; mov \%rdi,\%rdx; movzbl \%dh,\%edx; movzbl (\%rsi,\%rdx,1),\%ecx \\ \hline

    \multicolumn{1}{| c |}{\multirow {6}{*}{Rust-SGX SDK}}  & $<$dispose\_chunk$>$:0x8a  & mov 0x38(\%rsi),\%r9d; mov \%r9,\%rcx; lea (\%rdi,\%r9,8),\%r9; cmp 0x258(\%r9),\%rsi  \\ \cline{2-3}
    \multicolumn{1}{| c |}{}  &  $<$dispose\_chunk$>$:0x299 & mov 0x38(\%r8),\%r9d; mov \%r9,\%rcx; lea (\%rdi,\%r9,8),\%r9; cmp 0x258(\%r9),\%r8  \\ \cline{2-3} 
    \multicolumn{1}{| c |}{}  & $<$try\_realloc\_chunk.isra.2$>$:0x1eb  & mov 0x38(\%rsi),\%r9d; mov \%r9,\%rcx; lea (\%r12,\%r9,8),\%r9; cmp 0x258(\%r9),\%rsi  \\ \cline{2-3}
    \multicolumn{1}{| c |}{}  & $<$dlmalloc$>$:0x180b  & mov 0x38(\%rdx),\%r12d; mov \%r12,\%rcx; add \$0x4a,\%r12; cmp 0x8(\%rsi,\%r12,8),\%rdx  \\ \cline{2-3}
    \multicolumn{1}{| c |}{}   & $<$dlfree$>$:0x391  & mov 0x38(\%r8),\%edi; mov \%rdi,\%rcx; lea (\%rbx,\%rdi,8),\%rdi; cmp 0x258(\%rdi),\%r8  \\ \cline{2-3} 
    \multicolumn{1}{| c |}{}   & $<$dlfree$>$:0x467  & mov 0x38(\%rsi),\%edi; mov \%rdi,\%rcx; lea (\%rbx,\%rdi,8),\%rdi; cmp 0x258(\%rdi),\%rsi  \\ \hline 
	\end{tabular}
\caption{\attackName Type-II Gadgets in Popular SGX Runtimes.}
\label{table-gadgets2}
\end{center}
\end{table*}

\end{document}